\documentclass[twocolumn]{aastex631}

\newcommand{\acro}{\textsc{calamity}}
\newcommand{\Ninst}{N_\text{inst}}
\newcommand{\Ndata}{N_\text{data}}
\newcommand{\Nfreq}{N_\nu}
\newcommand{\Nant}{N_\text{ant}}
\newcommand{\psys}{\boldsymbol{u_\text{sys}}}
\newcommand{\pfg}{\boldsymbol{u_\text{fg}}}
\newcommand{\bgains}{\boldsymbol{g}}
\newcommand{\bvmodel}{\boldsymbol{m}}
\newcommand{\Nbl}{N_\text{b}}
\newcommand{\Nvec}{N_\text{v}}
\newcommand{\Nvecmax}{N_\text{v}^\text{max}}
\newcommand{\SBW}{\mathcal{W}}

\shorttitle{\acro}
\shortauthors{Ewall-Wice et al.}
\graphicspath{{./}{figures/}}
\usepackage{amsmath}

\begin{document}

\title{Precision Calibration of Radio Interferometers for 21\,cm Cosmology with No Redundancy and Little Knowledge of Antenna Beams and the Radio Sky. }

\correspondingauthor{Aaron Ewall-Wice}
\email{aaronew@berkeley.edu}

\author[0000-0002-0086-7363]{Aaron Ewall-Wice}
\affiliation{UC Berkeley Astronomy Department
 425 Campbell Hall
 University Dr., Berkeley 94720}

  \author[0000-0003-3336-9958]{Joshua S. Dillon}
 \affiliation{UC Berkeley Astronomy Department
 425 Campbell Hall
 University Dr., Berkeley 94720}
 
 \author[0000-0002-3240-9228]{Bharat Gehlot}
 \affiliation{School of Earth and Space Exploration
 Arizona State University Tempe, AZ}

\author[0000-0002-5400-8097
]{Aaron Parsons}
\affiliation{UC Berkeley Astronomy Department
 425 Campbell Hall
 University Dr., Berkeley 94720}
 
 \author{Tyler Cox}
 \affiliation{UC Berkeley Astronomy Department
 425 Campbell Hall
 University Dr., Berkeley 94720}

 \author[0000-0002-0917-2269]{Daniel C. Jacobs}
 \affiliation{School of Earth and Space Exploration
 Arizona State University Tempe, AZ}
 


\begin{abstract}
We introduce \acro{}, a precision bandpass calibration method for radio interferometry. \acro{} can solve for direction independent gains with arbitrary frequency structure to the high precision required for 21\,cm cosmology with minimal knowledge of foregrounds or antenna beams and does not require any degree of redundancy (repeated identical measurements of the same baseline). 
We have achieved this through two key innovations. Firstly, we model the foregrounds on each baseline independently using a flexible and highly efficient set of basis functions that have minimal overlap with 21\,cm modes and enforce spectral smoothness in the calibrated foregrounds. Secondly, we use an off-the-shelf GPU accelerated API ($\textsc{tensorflow}$) to solve for per-baseline foregrounds simultaneously with per-frequency antenna gains in a single optimization loop. GPU acceleration is critical for our technique to be able to solve for the large numbers of foreground and gain parameters simultaneously across all frequencies for an interferometer with $\gtrsim 10$ antennas in a reasonable amount of time. In this paper, we give an overview of our technique and using realistic simulations and demonstrate its performance in solving for and removing pathological gain structures to the level necessary to measure fluctuations in the 21\,cm emission field from Hydrogen gas during the Cosmic Dawn and Reionization.  If you want to start using \acro{} now, you can find a tutorial notebook at \url{https://github.com/aewallwi/calamity/blob/main/examples/Calamity_Tutorial.ipynb}
\end{abstract}

\keywords{}


\section{Introduction} \label{sec:intro}

Interferometry experiments are attempting to detect faint brightness temperature fluctuations in 21\,cm emission from Hydrogen in the early universe (see \citet{Liu:2020} for a review). Successful observations of these fluctuation will open an unrivaled window our Universe's Cosmic Dawn and Dark ages along with the evolution of large scale structure from shortly after recombination to the present day. Experiments targeting 21\,cm fluctuations include CHIME \citep{Bandura:2014}, Tianlai \citep{Chen:2015}, OWFA \citep{Subrahmanya:2017}, HIRAX \citep{Saliwanchik:2018}, the MWA \citep{Tingay:2013}, LOFAR \citep{VanHaarlem:2013}, NenuFAR \citep{Zarka:2012}, the LWA \citep{Eastwood:2019}, GBT \citep{Anderson:2018}, and HERA \citep{DeBoer:2017}.

The primary obstacle to a successful measurement of 21\,cm fluctuations are bright radio foregrounds which dwarf the cosmological signal by four-to-five orders of magnitude. Because they are spectrally smooth, these foregrounds can be circumvented by taking advantage of the fact that they are described by a small fraction of the spectral shapes that comprise 21\,cm fluctuations \citep{DiMatteo:2004, Morales:2004}. More precisely, foregrounds as observed by an interferometer are contained within a ``wedge'' shaped region of the 3D Fourier space of the sky \citep{Datta:2010, Morales:2012, Parsons:2012, Vedantham:2012, Pober:2013, Thyagarajan:2013}. In principal one can excise these wedge modes, leaving enough SNR for a detection \citep{Pober:2014}. 

While the intrinsic properties of foregrounds are amenable to 21\,cm signal recovery, spectral variations within direction dependent and direction independent antenna gains imprint additional spectral structures beyond the wedge. Sources of spectral structure that can leak power outside of the wedge include reflections sourced by impedance mismatches within the analog signal chain \citep{Beardsley:2016, EwallWice:2016b, Kern:2020}, direction dependent mutual coupling from reflections of sky signal between antenna elements \citep{Kern:2020, Fagnoni:2021, Josaitis:2021}, digital effects \citep{Prabu:2015, Barry:2019b}, injudicious choices in RFI flagging \citep{Offringa:2019} and polarization leakage \citep{Moore:2013, Kohn:2016, Asad:2016, Asad:2018}. Since the amplitude of foregrounds is roughly four-to-five orders of magnitude greater then the 21\,cm signal, leakage introduced by these effects must either be suppressed by a factor of roughly five orders of magnitude in the design of the instrument, or corrected to a similar level through calibration \citep{Trott:2016, Thyagarajan:2016, EwallWice:2016b, Patra:2018}.

To date, bandpass calibration methods either rely heavily on the correctness of a sky-model and per-antenna beams or make assumptions about the reproducibility of the beams and the exact positions of the antennas. Neither of the requirements under-girding calibration techniques have been realized in the field and it is unclear whether they can be. In this paper, we introduce a new approach that only relies on the more robust assumption that foregrounds, as observed by our antenna beams, occupy a limited number of smooth spectral modes contained within the wedge. Our technique does not require that the beams are identical between antennas nor does it require any degree of redundancy between baselines. It works well on both minimally redundant and redundant arrays. Our technique achieves this by solving for per-frequency gains across all spectral channels simultaneously with smooth foreground parameters on each individual baseline in the array. Our per-baseline smooth degrees of freedom approach stands in contrast to the existing techniques that attempt to model foreground emission and beam variations as sources with associated positions on the sky. For modern interferometers, this necessarily involves optimization of tens to hundreds of thousands of free parameters which we accomplish by implementing our technique using the off-the-shelf GPU accelerated $\textsc{tensorflow}$ API. We call our technique CALibration AMITY (\acro{}) because we believe it represents a significant improvement in the amity between detecting 21\,cm fluctuations and living with the direction independent gain features that arise in radio interferometers.

This paper is organized as follows. In \S~\ref{sec:BACKGROUND}, we give a short overview of calibration techniques being employed to calibrate 21\,cm cosmology arrays and briefly introduce our notation. In \S~\ref{sec:METHOD}, we discuss our modeling approach in formal detail along with how its performance scales with the number of antennas and frequency channels. In \S~\ref{sec:SIMULATIONS}, we discuss simulations of existing arrays to furnish several test-cases for our technique. In \S~\ref{sec:REDUNDANT}, we explore some of the differences to expect when applying \acro{} to a highly redundant versus a minimally redundant array (it can be used effectively on both). In \S~\ref{sec:LOWSNR}, we investigate whether \acro{}  In \S~\ref{sec:COMPUTATION} we discuss \acro{}'s runtime and memory scaling with array size. In \S~\ref{sec:CAVEATS}, we emphasize our techniques limitations and conclude in \S~\ref{sec:CONCLUSION}

\section{Previous Approaches to Calibration}\label{sec:BACKGROUND}
Calibration seeks to solve for and correct for the impact of $\Ninst$ different complex instrumental parameters given some set of $\Ndata$ interferometric measurements. Adopting the notation of \citet{Byrne:2021}, one typically models the observed cross-correlations between antenna $a$ and $b$ at frequency $\nu$, $\zeta_{ab\nu}$ as
\begin{equation}\label{eq:GENERALMODEL}
    \zeta_{ab\nu} = g_{a \nu} g^*_{b \nu} m_{ab\nu}(\psys, \pfg)
\end{equation}
where $g_{a\nu}$ are the direction-independent gain parameters for antenna $a$ at frequency $\nu$. If there are $\Nant$ antennas and $\Nfreq$ frequency channels, these gains form a set of $\Nfreq \Nant$ complex numbers. $m_{ab\nu}(\psys, \pfg)$ represents a model of all direction dependent systematics which includes intrinsic foregrounds which we parameterize with $\pfg$, and direction dependent systematics like beam effects and ionospheric distortions which we parameterize with $\psys$.

A number of strategies have been explored to solve for gains which broadly fall into two categories. Within each category, these techniques are distinguishable primarily in their approaches on how they parameterize $\bgains$ and $\bvmodel$, which parameters they solve for, and which they keep fixed.

    \subsection{Solve for Direction Independent Gains.} In this approach, one assumes a fixed model for the foregrounds and the direction dependent antenna beams and attempts to solve for $\bgains$ only. This method was an excellent strategy for instruments whose fields of view could be dominated by a single bright and well understood source (e.g. \citealt{Baars:1965}) and the conservative approach of not allowing the sky to vary can reduce the risk of signal loss. Direction independent gain solvers being used for 21\,cm cosmology include $\textsc{stefcal}$ \citep{Salvini:2014} and $\textsc{cubical}$ \citep{Sob:2020}, self-holography \citep{Gueuning:2020}.
    The large fields of view at low frequency and the extreme dynamic range requirements of $\sim10^{5}$ pose a serious obstacle since any calibration observation will have a multitude of sources off of boresight contributing significantly to observed visibilities. Hence, $\bvmodel$ must not only involve accurate models of faint sources which are today highly uncertain \citep{Jacobs:2013} but also precision a-priori knowledge of the beam(s). The state-of-the-art in beam modeling today is far behind the level of precision required for 21\,cm \citep{Pober:2012, Newburgh:2014, Neben:2015, Colgate:2015, Sutinjo:2015, Berger:2016, Neben:2016, Sokolowski:2017, Jacobs:2017, Line:2018, deLeraAcedo:2018, Nunhokee:2020}. If we fit for enough spectral degrees of freedom to correct intrinsic gain structures outside of the wedge, beam and source model errors leak power into the EoR window at levels that far exceed the 21\,cm signal \citep{Barry:2016, Trott:2016, EwallWice:2017, Joseph:2020}. While this might be circumvented by excluding / downweighting long baselines \citep{EwallWice:2017}, doing so requires an accurate model of diffuse emission which is highly polarized \citep{Lenc:2016}. Polarized diffuse sky models are being constructed for calibration \citet{Byrne:2021b} and work remains on determining the requirements on such a model for short baseline calibration to be effective.  Even if the beams and foregrounds were known to sufficient accuracy, ionospheric fluctuations corrupt calibration solutions (e.g. \citealt{Intema:2009, Jordan:2017, Gehlot:2018, Yoshiura:2021}) albeit in a way that may average down with time \citep{Vedantham:2015}. Thus, solving for direction independent gains given a sky-model has so-far required making potentially innacurate a-priori assumptions of the gains being spectrally smooth (e.g. \citealt{Beardsley:2016, Mertens:2020}) while fine-frequency degrees of freedom must be modeled using other measurements such as autocorrelations (e.g. \citealt{EwallWice:2016b, Li:2019, Kern:2020b}), beam variations \citep{Line:2018}, and simulations of digital effects \citep{Barry:2019b}. In its current state, gain-only calibration requires spectrally smooth priors which may not be realized in the field.
    
    \subsection{Solve for Direction Dependent Gains and a model of the Sky.}
   One solution to resolving beam, ionosphere, and foreground modeling errors is to explicitly solve for them. Foreground models are often encoded as a list of source locations and spectral parameters but have also been represented by extended spatial shapes such as wavelets \citep{Gu:2013, Chapman:2013} to more efficiently describe diffuse emission. Packages such as $\textsc{ddfacet}$ \citep{Tasse:2018}, $\textsc{sagecal}$ \citep{Kazemi:2011, Kazemi:2013}, the MWA Realtime System \citep{Mitchell:2008}, $\textsc{facetcal}$ \citep{VanWeeren:2016} and $\textsc{cubical}$ \citep{Kenyon:2018, Sob:2020} all attempt to solve for the direction dependent Jones matrices of the antennas along with a parameterized source model (such as locations and spectral evolution) and ionospheric distortions. Several of these methods such as the $\textsc{rts}$ and $\textsc{sagecal}$ also have direction independent functionality which has appeared in the literature. The primary obstacle to increasing the number of degrees of freedom in our foreground and telescope models is the prospect of unintentionally subtracting components of the 21\,cm signal itself. For example \citet{Patil:2016} report potentially problematic suppression of diffuse polarized emission by $\textsc{sagecal}$. The solution these authors adopt in \citet{Patil:2017} is to exclude short baselines (which are dominated by the diffuse emission) from calibration but this runs into the issue of contamination from long baselines if the beams are not identical between antennas. Further refinements to algorithms such as $\textsc{sagecal-co}$ \citep{Yatawatta:2016, Gehlot:2019, Mertens:2020} use concensus optimization to force the direction dependent gains to follow smooth polynomials. This mitigates the contamination from long baselines along with signal suppression but still leaves unresolved the issue of fine frequency scale direction independent gains.
   
A popular method for constraining a sky and beam model with arbitrary frequency resolution while attempting to limit signal loss and errors incurred by innacuracies in apriori models is to assume that all antenna beams are identical and exploit the redundancy between repeated copies of the same baseline to solve for sky parameters and gains up to some frequency dependent degeneracies before referring back to the sky \citep{Wieringa:1992, Liu:2010, Zheng:2013, Zheng:2014, Dillon:2020}. Unfortunately, high redundancy is difficult to accomplish due to mechanical variations between antennas \citep{Line:2018, Kim:2021}, positional errors, and over-the-air mutual coupling between antennas \citep{Fagnoni:2021, Josaitis:2021}. \citet{Joseph:2018} demonstrate that small errors in antenna positions lead to significant biases in redundant calibration solutions and \citet{Orosz:2019} find that gain solution errors arising from expected levels of non-redundancy introduce foreground contamination can exceed 21\,cm fluctuations by orders of magnitude while \citet{Choudhuri:2021} determined that these errors also cause mild decoherence in the 21\,cm fluctuations themselves. The equations in redundant calibration still admit frequency dependent degeneracies which must still be solved for by referencing back to a sky-model and the associated errors can be amplified by redundant antenna arrangements \citep{Byrne:2019}.

   Several authors have proposed ``hybrid'' approaches such as solving for non-redundant degrees of freedom \citep{Sievers:2017}, projecting the degeneracies in a redundant sub-array calibration solution onto a sky-model based calibration solution \citep{Li:2018}, and using sky-models to inform priors on redundant solutions \citep{Byrne:2021}. Some of these techniques demonstrate improvements over purely redundant or sky-based approaches. None have demonstrated that the errors in gain solutions with arbitrary spectral degrees of freedom are suppressed sufficiently in the presence of non-redundancies to allow for a 21\,cm detection.

All calibration approaches must balance a tradeoff between flexibility and signal loss. A calibration technique must be flexible enough to model and correct fine-scale frequency structures in the instrumental gains along with antenna-to-antenna variations in the beams. However, more flexible techniques run the risk of subtracting actual 21\,cm fluctuations or introducing additional errors. An example of the latter, calibration with a fixed sky-model but with completely flexible frequency degrees of freedom in the gains compensates for the disagreement between the true and measured sky on long baselines by inserting fine-scale frequency structure. When performed on an array that is not truely redundant, redundant calibration does the same thing. Solving for the direction dependent gains towards bright foregrounds as point sources adds additional flexibility but still introduces errors since the radio sky is comprised of large numbers of faint and extended sources that are not completely encapsulated by any point source model. For example, \citet{Yoshiura:2021}, find that solving for direction dependent gains in the directions of a small number of the brightest sources actually increases systematics in the EoR window over direction independent calibration.

In this paper, we introduce \acro{} which is a calibration method that is flexible enough to deal with non-redundant beams and direction dependent effects, and a completely unconstrained radio sky while at the same time can be demonstrated to not introduce biases within the EoR window that can prevent a robust 21\,cm detection. Most existing calibration techniques model direction dependent effects as a source model tied to the sky modulated by different direction dependent systematics. Our approach is to instead model both the direction dependent effects and the foregrounds per-baseline as linear combinations of discrete prolate spheroidal sequences which are a maximally efficient basis for modeling band limited signals \citep{Slepian:1978} and have proven to be excellent for modeling per-baseline foregrounds \citep{EwallWice:2021}. More explicitly, we model the visibility formed from antennas $a$ and $b$, including all direction dependent beam effects, $m_{ab\nu}$ from equation~\ref{eq:GENERALMODEL} as 
\begin{equation}\label{eq:FGMODEL}
   m_{ab\nu} = \sum_i A_{\nu i}(\SBW_{ab}) u_{iab} 
\end{equation}
where $A_{i\nu}(\tau_{ab})$ is the discrete prolate spheroidal sequence (DPSS) \citep{Slepian:1978} with a length of $\Nfreq$ and standardized half bandwidth of $\SBW_{ab} = 2 \Nfreq \Delta \nu \tau_{ab}$ where $\tau_{ab}$ is a delay approximately equal to the light travel time between antenna $a$ and antenna $b$ (the horizon delay at the edge of the wedge) and $\Delta \nu$ is the channel width. Each $u_{iab}$ is a coefficient which must be solved for by fitting to measurements and each visibility has its own series of $u_{iab}$. The number of DPSS vectors required to describe band-limited signals with delay half-width of $\approx \tau_{ab}$ is $\Nvec(\tau_{ab}) \approx \SBW_{ab}$ \citep{Slepian:1978}. Thus, we model each intrinsic visibility with a sum of one-dimensional sequences in frequency. We refer the reader to \citet{EwallWice:2021} for a more detailed description of this modeling strategy.

Doing this, instead of sky-based modeling, ignores correlations between the baselines and limits any increases in variance and signal loss, to the foreground wedge while achieving a high degree of flexibility to model systematics. Our method's philosophy is to model the foregrounds and direction dependent effects with a highly flexible and degenerate set of parameters which incur significant variance and biases within the wedge while leaving the EoR window relatively free of biases.

We implement our technique in $\textsc{python}$ using the optimization and auto-differentiation machinery of the $\textsc{tensorflow}$ library \citep{Tensorflow:2015} which provides a readable and flexible code base for large-scale modeling and optimization problems with out-of-the-box GPU support. 





\section{\acro}\label{sec:METHOD}
In this section, we describe \acro{} in detail \S~\ref{ssec:DESCRIPTION}, how it can be modified to reproduce the redundant calibration approach \S~\ref{ssec:REDUNDANTFORMALISM} (though we can calibrate redundant arrays perfectly well without this modification).
\subsection{A Description of the Method}\label{ssec:DESCRIPTION}

Incorporating equation~\ref{eq:FGMODEL} into equation~\ref{eq:GENERALMODEL} yields the complete model for our data as
\begin{equation}
   \zeta_{ab\nu} = g_{a\nu} g_{b\nu}^* \sum_i A_{ab\nu, i}(\SBW_{ab}) u_{i,ab}
\end{equation}

Since we model each visibility with an independent set of $u_{iab}$, our model is flexible enough to account for all variations caused by the ionosphere and non-redundant beams as long as they are below $\tau_{ab}$ for a particular baseline. 

To take advantage of fast rectilinear matrix operations on a GPU we represent our foreground and direction dependent systemics parameters, $\boldsymbol{u}$, as an $\Nbl \times \Nvecmax$ tensor where $\Nvecmax$ is the maximum number of vectors needed to model any of our baselines and is approximately equal to the maximum standardized half bandwidth of any baseline in our array $\text{Max}_{\sim(a,b)}(\SBW_{ab})$. Because many of our short baselines require fewer components to model, this means that a large fraction of our $\boldsymbol{u}$ elements are zero. We represent $\boldsymbol{A}$ as an $\Nvecmax \times \Nbl \times \Nfreq$ tensor which dominates our technique's memory requirements. We set the rows of $\boldsymbol{A}$ that correspond to modeling vectors that are unused on short baselines to zero as well.

To solve for $\boldsymbol{u}$ and $\boldsymbol{g}$, we minimize the mean-squared errors between our instrument model and the measured visibilities  $v_{ab\nu}$,
\begin{equation}\label{eq:LOGLIKELIHOOD}
\mathcal{L}(\boldsymbol{u}, \boldsymbol{g}) = \sum_{ab\nu} w_{ab\nu} \left| v_{ab\nu} - g_{a\nu} g_{b\nu}^* \sum_i u_{abi} A_{iab\nu}  \right|^2,
\end{equation}
using the $\textsc{adamax}$ algorithm \citep{Kingma:2014}. $w_{ab\nu}$ stands for a per-baseline and per-frequency weight which can take into account, for example, flagged channels and differing noise levels per-baseline. While it is computationally expedient, first order gradient descent will only converge to the local minimum closest to our initial guesses for $\boldsymbol{g}$ and $\boldsymbol{u}$. Thus, our technique as written should be used as a second stage of calibration after a first stage that obtains initial guesses that fall within the convex region containing the global minimum of equation~\ref{eq:LOGLIKELIHOOD} such as $\textsc{firstcal}$ \citep{Parsons:2014, Dillon:2018, Li:2018, Dillon:2020}. As written, the cost function in equation~\ref{eq:LOGLIKELIHOOD} is completely degenerate with respect to a single complex gain multiplier. We break this degeneracy by adding a minimal prior, $p(\boldsymbol{u})$ to $\mathcal{L}$ on the sum over all baselines and channels in the data.
\begin{equation}
    p(\boldsymbol{u}) = \left| \sum_{ab\nu} w_{ab\nu}^\prime \left( \sum_i \left(u_{abi} A_{iab\nu}\right) - m_{ab\nu}\right)\right|^2
\end{equation}
where $m_{ab\nu}$ is a sky model which can have relatively low detail and merely serves to fix degeneracies like the  overall gain scale. $w_{ab\nu}^\prime$ is a set of prior weights which we can adjust to set the strength of the prior. For this work we set $w^\prime_{ab\nu} = w_{ab\nu} = (\Nbl \Nfreq)^{-1}$ for all weights. One could also drop this prior from the optimization since the degeneracy, by definition does not effect the first likelihood term and set an overall gain scale after completing optimization by taking the ratio between the sum of measured visibilities and the sum of model visibilities.

\subsection{Redundant Calibration with \acro{}}\label{ssec:REDUNDANTFORMALISM}
If we are highly confident that our visibilities are actually redundant, we can reduce the set of modeling vectors $u_{abi}$ and $A_{iab\nu}$ coefficients so that a single set of modeling vectors and coefficients describes all baselines within a redundant group. If we label our redundant groups by $R$, then equation~\ref{eq:LOGLIKELIHOOD} reads
\begin{equation}\label{eq:LOGLIKELIHOODREDUNDANT}
\mathcal{L}_\text{red}(\boldsymbol{u}, \boldsymbol{g}) = \sum_R \sum_{(a,b) \in R, \nu} w_{ab\nu} \left| v_{ab\nu} - g_{a\nu} g_{b\nu}^* \sum_i u_{Ri} A_{iR\nu}  \right|^2.
\end{equation}
Thus, \acro{} is equipped to perform standard redundant calibration if the user so wishes.






\section{The Performance of \acro{} in Recovering 21\,cm Fluctuations in the EoR window in Realistic Simulations}\label{sec:SIMULATIONS}

In this section, we test \acro{} on simulated visibilites for a minimally redundant array containing realistic foregrounds and calibration errors. Since \acro{} can calibration a minimally redundant array with minimal sky knowledge, it follows that it can calibrate nearly redundant arrays (redundant arrays with positional and beam errors) as well. We will briefly address the redundant case in \S~\ref{sec:REDUNDANT}  In \S~\ref{ssec:SIMULATION}, we describe our simulations

\subsection{The Simulations}\label{ssec:SIMULATION}
We test \acro{} on examples of two different array paradigms in wide use today -- a highly redundant hexagonal layout and a minimally-redundant layout. Beyond the already discussed calibration strategy that they enable, redundant arrays are intended to enhance sensitivity to a handful of modes and increase our leverage on systematics by including many repeated copies of the same baseline separation. This  allows us to beat down per-antenna features by crossing and averaging different copies of the same redundant baseline \citep{Parsons:2014, Ali:2015, HERA:2021}. Randomized arrays have the advantage of reducing the risk of signal loss from calibration overfitting by averaging more independent cosmological modes into each k-bin. \citet{Lanman:2019} also note that this reduces sample variance. We compare \acro{}'s performance on both types of arrays by considering a minimally-redundant and redundant subset of antennas from the compact configuration of the MWA phase-II \citep{Wayth:2018}. For our minimally-redundant array, we choose the first thirty-six antennas that comprise the majority of the MWA's compact minimally-redundant core. For our redundant array, we use one of the MWA phase-II's two thirty-six antenna hexagonal patterns. The primary beam model for MWA we used for the visibility simulations is described in \cite{Sutinjo:2015} and \cite{Sokolowski:2017}. Our simulations are run using the $\textsc{pyuvsim}$ visibility simulator package\footnote{\url{https://github.com/RadioAstronomySoftwareGroup/pyuvsim}} \citep{Lanman:2019b}. We simulate fifty-six, two second integrations starting at an LST of $\approx 1.8$ hours. Visibilites are simulated for 384~channels with 80~kHz channel width spanning $167-187$~MHz bandwidth. The foreground model for the simulated visibilities is taken from the GLEAM catalog \citep{Hurley-Walker:2016} contains all sources with wide-band integrated flux $F_{\text{int}}>50$~mJy. Sources in the foreground model are assumed to be flat-spectrum over the simulation bandwidth, and the a single wide-band integrated flux value per source is used for all frequency channels. While the true foregrounds are not spectrally flat, the majority of spectral variation originates from antenna beams anyways and \citet{EwallWice:2021} have established the robustness of using DPSSs to model foregrounds with direction dependent spectral variation. The EoR model used for the simulations is a flat power spectrum $P(k) = 12482.356$~$\text{K}^2\text{Mpc}^3$. This corresponds to noise sky of variance $9$~$\text{K}^2$ in image space in zeroth channel. The EoR model is generated in the healpix projection with nside=256 that corresponds to a spatial resolution of $13.74$~arcmin. The EoR model is stored in a \texttt{SkyModel} object of \textsc{pyradiosky}\footnote{\url{https://github.com/RadioAstronomySoftwareGroup/pyradiosky}} package for seamless readability by \textsc{pyuvsim}. Table~\ref{tab:simulation-details} summarizes the simulation details.

\begin{table*}
\centering
\caption{Visibility simulation details.}
\label{tab:simulation-details}
\begin{tabular}{ll}
     \hline
     Parameter & Value \\
     \hline
     Simulation Software    & \textsc{pyuvsim} \\
     Instrument layout      & MWA Phase-II (Spiral and two hexagonal sectors) \\
     Number of Antennas ($N_{\text{ant}}$) & 128  \\
     Primary Beam & MWA embedded element beam model \citep{Sutinjo:2015,Sokolowski:2017}\\
     Frequency range & 167-187~MHz\\ 
     Number of channels ($N_{\text{ch}}$) & 384 \\ 
     Frequency resolution ($\Delta \nu$) & 80~kHz \\
     Number of timestamps ($N_{\text{t}}$) & 56 \\
     Temporal resolution ($\Delta t$) & 2.0~s\\
     Sky-models & GLEAM ($F_{\text{int}}>50$~mJy, $N_{\text{src}}\sim 2.83\times 10^5$)\\
                & Flat-spectrum EoR ($P(k) = 12482.356$~$\text{K}^2\text{Mpc}^3$, $N_{\text{src}}\sim 7.86\times 10^5$)\\
    \hline
\end{tabular}
\end{table*}

\begin{figure*}
    \centering
    \includegraphics[width=\textwidth]{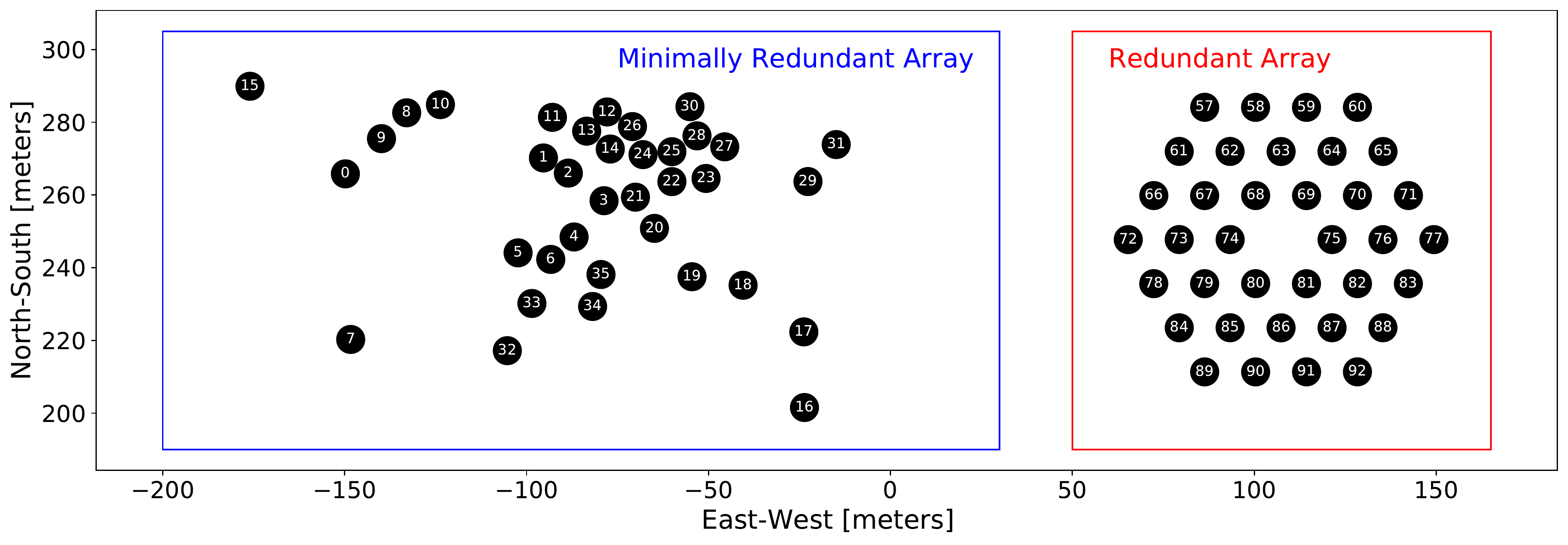}
    \caption{We test \acro{} on simulations of two sub-arrays of the MWA. A minimally-redundant array consisting of antennas 0 through 35 (left) and one of the MWA's redundant hexagonal sectors (right).}
    \label{fig:LAYOUTS}
\end{figure*}

\subsubsection{Reduction of Simulation Products and Power Spectra}
Our foreground simulation contains beam interpolation artifacts which set an intrinsic systematics floor at $\approx -45$\,dB below the foreground brightness levels, therefor we scale our EoR fluctuations to be $\approx -40$\,dB below the level of foregrounds. While this is brighter then the expected EoR signal levels, this arbitrary scaling does not affect the conclusions of this paper. For each baseline and time, we simulate two independent realizations of thermal noise scaled to have a standard deviation on each baseline that is $10\times$ the level of the 21\,cm signal and generate two effective measurements with an independent noise realization at each baseline, frequency and time which we will interleave to form power spectra in the fashion of \citet{Dillon:2014}. We set our simulated gains to be constant over the 112 seconds of observation. We set each frequency and time sample to be unity plus an independent normally distributed real and imaginary component with an amplitude of $0.1$. Our simulated visibilities can thus be broken down into
\begin{equation}
    V_{ab}(\nu, t) = g_a(\nu) g_b^*(\nu) \left[ F_{ab}(\nu, t) + S_{ab}(\nu, t) + N_{ab}(\nu, t) \right]
\end{equation}
where $F_{ab}$ is the foreground visibility for baseline $ab$, $S_{ab}$ is the EoR visibility and $N_{ab}$ are the even (odd) noise realizations. To demonstrate \acro{}'s ability to calibrate arbitrarily fine gain structures, we set our frequency dependent simulated gains to be highly chromatic and equal to a normally distributed noise with a standard deviation in the real and imaginary parts of 0.1 added to a baseline gain of unity. We assume that the gains are constant in time.

Running \acro{} provides us with estimates of gains and foregrounds for each even/odd integration which we denote as $\widehat{g}_a(\nu)$ and $\widehat{F}_{ab}(\nu, t)$ respectively. We compute a calibrated residual at each time and frequency,
\begin{equation}
    R_{ab}(\nu, t) = \left[ \hat{g}_a(\nu, t) \hat{g}^*_b(\nu, t) \right]^{-1} \left[V_{ab}(\nu, t) - F_{ab}(\nu, t)\right].
\end{equation}

We fringe-stop each residual and average coherently over the full 112 seconds of observation. This has been demonstrated to lead to negligible signal loss for the baseline lengths under consideration \citep{Ali:2015}. We next form power spectra for pairs of baselines $ab$ and $cd$ at two different times $t$ and $t^\prime$ by computing the complex conjugated products between delay-transformed simulated data with our two independent noise realizations and scaling to cosmological power spectrum units using the relation \citep{Parsons:2012}
\begin{equation}
    P_{ab, cd}^{V}(\boldsymbol{k_\perp}, k_\parallel, t, t^\prime)
    = \frac{X^2Y}{\Omega_{pp}B_{pp}}\widetilde{V}_{ab}(\tau, t) \widetilde{V}^{*}_{cd}(\tau, t^\prime)
\end{equation}
where $\boldsymbol{k_\perp}=2 \pi {\bf b}_{ab} / X, k_\parallel = 2 \pi \tau / Y$, $\Omega_{pp}$ is the solid angle integral of the primary beam squared, $B_{pp}$ is the frequency integral of the passband squared, $X$ and $Y$ are linear scaling factors relating transverse and parallel distances to angle and frequency distances respectively and $\widetilde{V}_{ab}(\tau, t)$ is the delay-transform of $V_{ab}(\nu, t)$ \citep{Parsons:2012a}
\begin{equation}
    \widetilde{V}_{ab}(\tau, t) = \Delta \nu \sum_\nu e^{2 \pi I \tau \nu} V_{ab}(\nu, t) T(\nu),
\end{equation}
where $T(\nu)$ is a frequency tapering function and $\Delta \nu$ is the channel width. For all power spectra, we use a Blackman-Harris taper. We compute power spectra for the uncalibrated visibilities ($P^{V}$) as well as our calibrated residuals ($P^{R}$), the EoR fluctuations ($P^{S}$), noise only ($P^{N}$) and the EoR fluctuations combined with noise and foregrounds ($P^{S+F+N}$).
For the redundant array, we only form products between different baselines within the same redundant group ($ab \neq cd$). For the minimally redundant array, we take the products of baselines with themselves ($ab = cd$). We estimate error bars based on the RMS amplitude of thermal noise power spectrum, $\sigma_N$ and the RMS amplitude of the 21\,cm power spectrum $\sigma_S$ \citep{McQuinn:2006},
\begin{equation}
    \sigma_P^2 = \left(\sigma_N + \sigma_s\right)^2.
\end{equation}
When we calculate noise only power spectra, we neglect the $\sigma_S$ term and when we calculate signal only power spectra we neglect the $\sigma_N$ terms.





\subsection{Simulation Results}
We now discuss the performance of \acro{} in calibrating a minimally redundant array. Before diving into the detailed results, it is illustrative to inspect how well our fitted gains and calibrated data match the injected gains and data to better understand where we can trust \acro{}.

\begin{figure*}
    \centering
    \includegraphics[width=\textwidth]{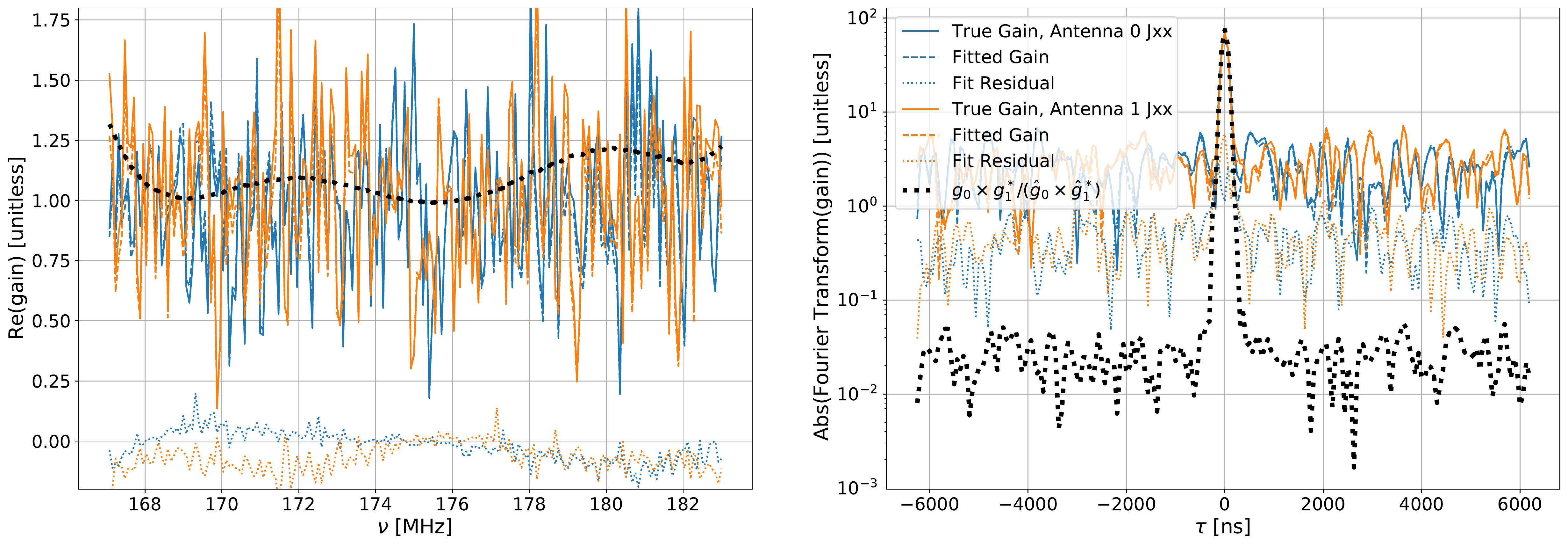}
    \caption{A comparison between the real parts of our simulated (solid lines) and recovered (dashed lines) gains for antennas 0 and 1, part of our ``Minimally Redundant'' array. We use gains that are completely uncorrelated between frequencies to make the spectral calibration challenge as difficult as possible. (This is substantially more chromatic then what is typically encountered in the field.) Our gain estimates do not agree particularly well with the true injected gains. In fact, their residuals are on the order of 10\% (dotted lines). {\bf Right Panel:} Even in the delay domain (with a Blackman-Harris taper), the residuals are at 10\% at all delays. While it may be tempting to think that such large residuals have catastrophic implications for 21\,cm cosmology, they arise from degeneracies between the foreground wedge and the gains and between the different gain frequencies themselves. The gain correction ratio, $g_0 \times g_1 / (\hat{g}_0 \times \hat{g}_1^*) - 1$ (black dotted line) is smooth at the level of the thermal noise fluctuations. Thus, the calibrated foregrounds agree with our data at high precision (Fig.~\ref{fig:CALIBRATED_RESID}).   }
    \label{fig:GAIN_RESID}
\end{figure*}

\begin{figure*}
\includegraphics[width=\textwidth]{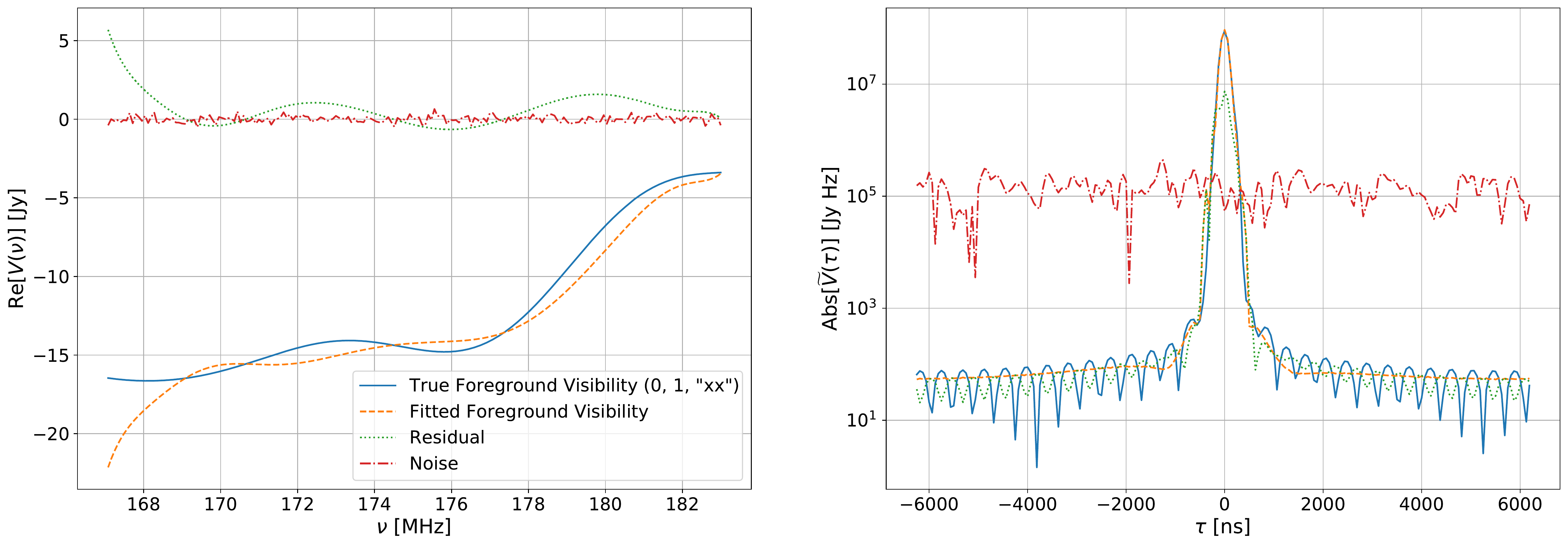}
\caption{{\bf Left Panel:} Comparison between the true simulated foregrounds (dashed lines) and our fitted foreground model (dashed lines) on a single baseline and time snapshot. Like the gains (Fig.~\ref{fig:GAIN_RESID}), our foreground model does not agree particularly well with the True foregrounds and the residual (dotted line) is on the order of $\approx 10-20\%$ and significantly exceeds the level of the thermal noise. {\bf Right Panel:} The same as the left, but in delay space with a Blackman-Harris taper. By definition, our foreground model resides within a compact region of delay space. As the model is contained within the foreground wedge, subtracting it does not introduce any additional spectral structures beyond the wedge.}
\label{Fig:MODEL_RESID}
\end{figure*}

\begin{figure*}
    \centering
    \includegraphics[width=\textwidth]{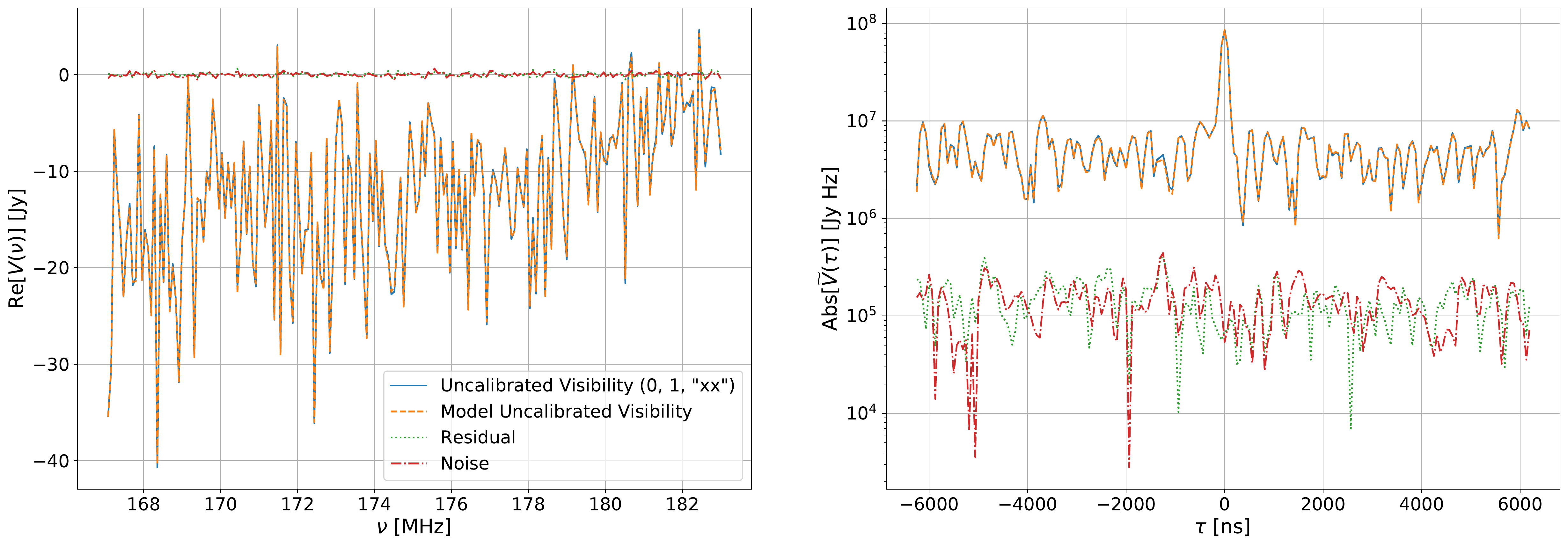}
    \caption{{\bf Left Panel:} Comparison between the real part of an uncalibrated visibility in our simulations at a single time snapshot (solid line) with our best fit model (dashed line). While our fitted gains are in relatively poor agreement with the true gains (Fig.~\ref{fig:GAIN_RESID}), our fitted model visibilities have excellent agreement and at the same level as the thermal noise (dash-dotted line). {\bf Right Panel:} The same comparison in delay space. We see that while fitted gains and foregrounds alone only agree at the $10\%$ level, our visibility model is accurate at the $0.1\%$ level and at a similar level to the simulated thermal noise at all delays. Thus, if we subtract our visibility model, we can eliminate the systematics caused by spectral bandpass features leaking power into high delays. For this reason, \acro{} might be better thought of as a precision filter rather then a precision calibration technique.}
    \label{fig:CALIBRATED_RESID}
\end{figure*}

We inspect the fitted gains for antennas $0$ and $1$ in our minimally redundant array in Fig.~\ref{fig:GAIN_RESID}. Our fitted gains to not agree particularly well with the true gains, only on the $\lesssim$ 20\% level on all delays. However, the quantity that governs whether a calibrated visibility formed from antennas $a$ and $b$ is smooth is $g_a \times g_b^* / (\hat{g}_a \times \hat{g}_b^*)$. In the right hand panel of Fig.~\ref{fig:GAIN_RESID}, this quantity is much smoothers. We see in Fig.~\ref{Fig:MODEL_RESID} that our foreground model on this baseline only agree at the $10\%$ level. The disagreements between the foregrounds and gains alone arise from frequency dependent degeneracies which might be broken by incorporating additional information such as baseline-baseline covariances or a more stringent prior on the foreground model and beams, neither of which are available with the current state-of-the art. While we may not have have high-precision estimates of the foregrounds or the gains alone, we do obtain high precision estimates of the actual quantity we optimized for by minimizing equation~\ref{eq:LOGLIKELIHOOD} -- the products of the gains with the foreground model. In Fig.~\ref{fig:CALIBRATED_RESID} we show our model for the measured visibility between antennas $0$ and $1$ which is the product of the two gains and the model for that visibility. While the foreground model and gains alone only agree with the true gains and foregrounds at the $10\%$ level, the products of the gains and foregrounds are in excellent agreement and are identical to within the uncertainties from thermal noise. If we subtract the model of foregrounds leaked by gain errors. If we subtract our estimate for the foregrounds multiplied by the gains, then we eliminate the contamination introduced by foregrounds being leaked by gains into fine-spectral scales. This is precisely what \acro{} does and in this sense, it might be better thought of as a filter then a calibration technique since it identifies and subtracts power that is described by wedge-like foreground modes leaked by antenna gains at high precision. 

\begin{figure*}
    \centering
    \includegraphics[width=\textwidth]{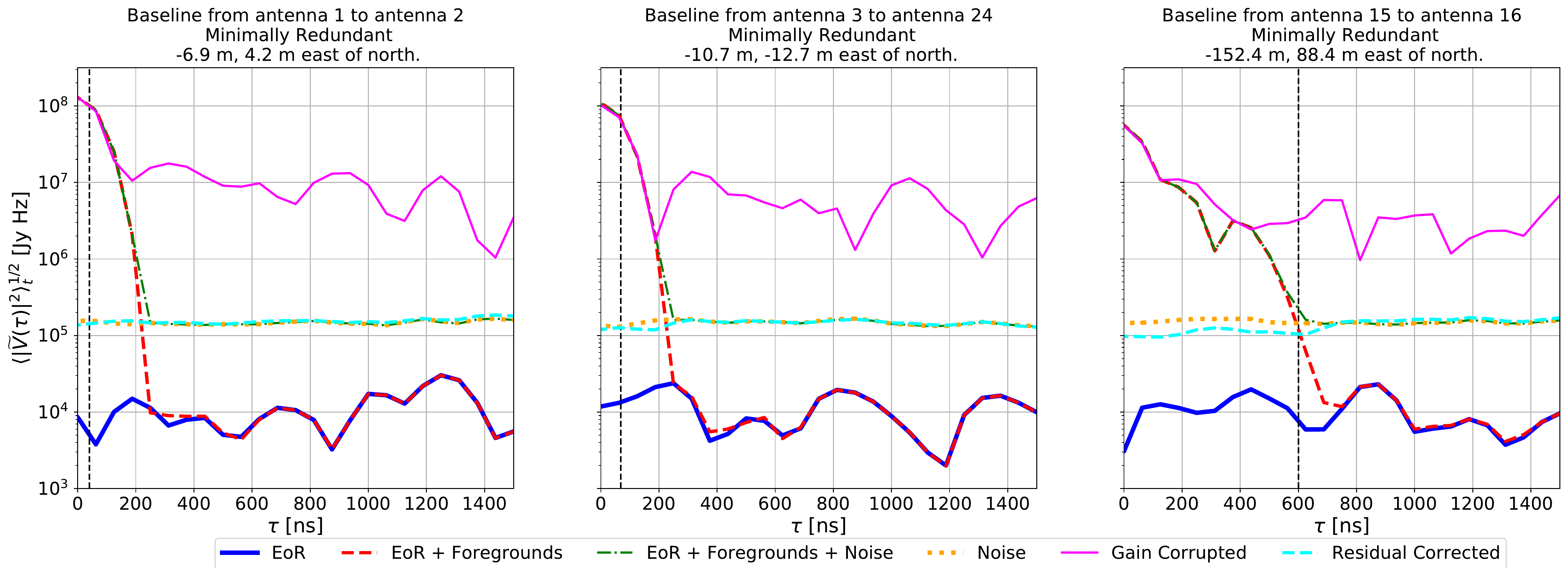}
    \caption{The RMS over time of simulated visibility amplitudes for several baselines in our minimally redundant layout. \acro{} solves for foregrounds multiplied by gain structures (thin solid magenta line) and subtracts the model to obtain residuals (dashed cyan line). Outside of the wedge (vertical black dashed line), the RMS amplitudes of our calibrated residuals (cyan dashed line) are reasonably consistent with thermal noise fluctuations (dotted orange line). Our per-baseline modeling approach cannot distinguish between foregrounds and noise / EoR fluctuations inside of the wedge, leading to signal loss which is observable here as a small decrease in residual amplitude. Coherent time averages and cross-multiplication between independent noise realizations lead to a more visible decrease (see Figs.~\ref{fig:WEDGE}~and~\ref{fig:INSPECT_BLS}).}
    \label{fig:INSPECT_BLS}
\end{figure*}

In Fig.~\ref{fig:INSPECT_BLS}, we inspect the RMS average of the delay-transformed visibilities. We see that the per-baseline $\widetilde{R}$ values recovered by \acro{} are broadly consistent with the levels of injected thermal noise fluctuations and orders of magnitude lower then the injected gain features.

While the residuals are broadly consistent with the noise outside of the wedge, there is a noticeable reduction in amplitude inside of the wedge. This is due to the fact that our per-baseline modeling cannot distinguish between foreground modes and thermal noise or cosmological fluctuations. Thus, subtracting our guess at the the foregrounds leads to signal loss inside of the wedge. What does remain within the wedge are errors in the gain solutions which exist at a lower level then the thermal noise. For the redundant array, we also find that we get roughly the same residual when calibrating with redundant modeling using a single set of foreground coefficients to describe all baselines within each redundant group (equation~\ref{eq:LOGLIKELIHOODREDUNDANT}). We hypothesize that this is because when we use multiple foreground coefficients within a truly redundant group, we are simply adding repeated copies of the same set of parameters which has no effect on our final answer.

\begin{figure*}
    \centering
    \includegraphics[width=\textwidth]{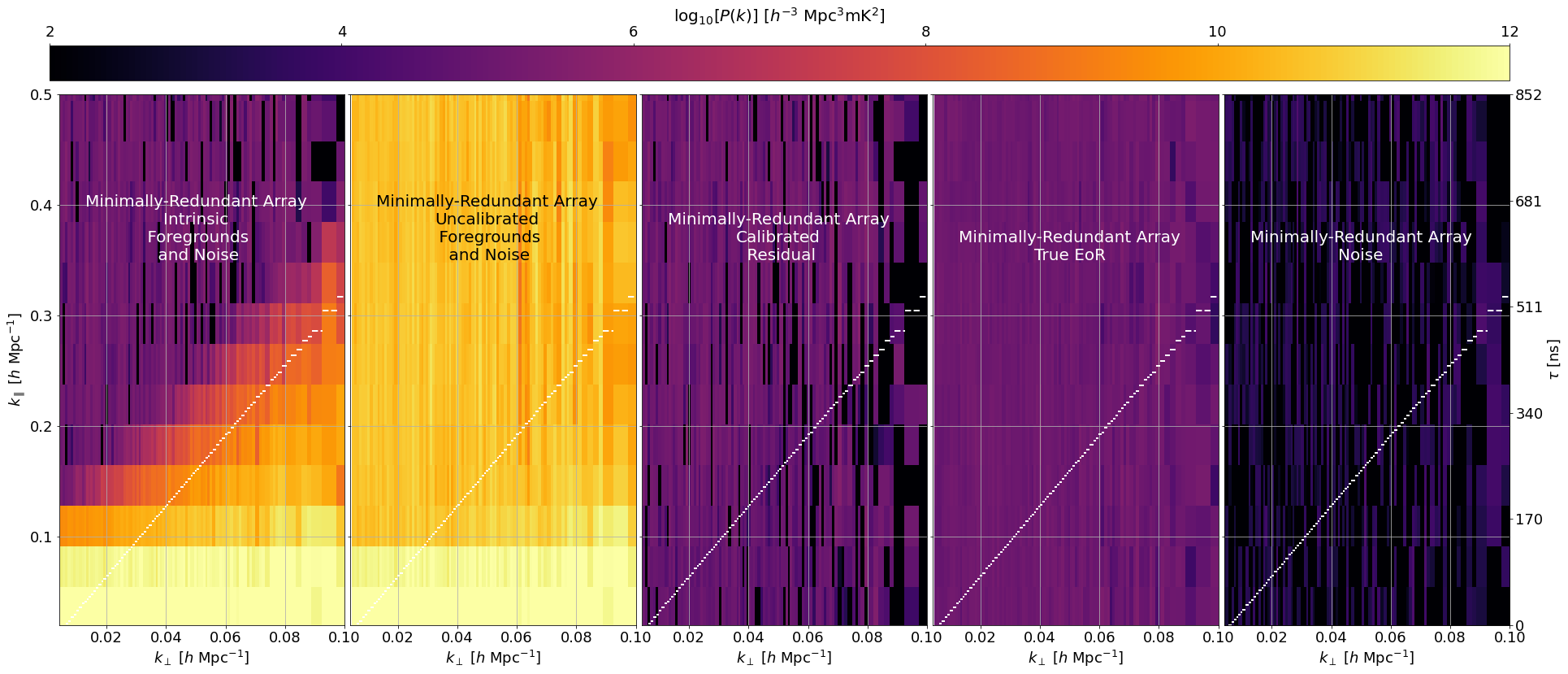}
    \caption{From left to right, cylindrically averaged power spectra for our simulated foregrounds, noise, and EoR model, $P^{S+N+F}$, uncalibrated visibilities, $P^{V}$,  calibrated residuals $P^{R}$, the injected EoR fluctuations, and thermal noise only.}
    \label{fig:WEDGE}
\end{figure*}

We inspect the real parts of wedge power spectra in Fig.~\ref{fig:WEDGE}. In our simulated visibilites without calibration errors, foreground power is contained within the wedge (left panel). Our calibration solutions are uncorrelated between frequencies and as a result, the entire EoR window is filled in with systematics (left of center panel). After running \acro{}, these systematics are effectively removed (center panel) and over most cylindrical modes, these residuals are roughly consistent with the injected EoR signal level (center right panel) albiet with clear postive and negative fluctuations arising from noise (right panel). To further beat down noise, we must average in spherical k-bins.

\begin{figure*}
\includegraphics[width=\textwidth]{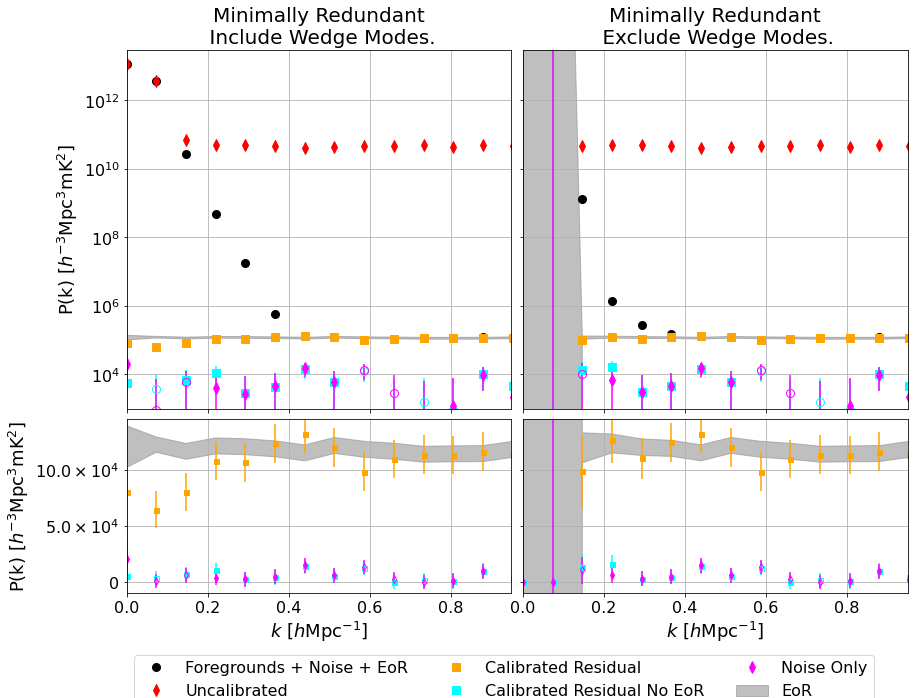}
\caption{{\bf Top Row:} Spherically averaged power spectra for our simulations of foregrounds, 21\,cm fluctuations, and noise~$P^{S+F+N}$ (black circles), the same simulations corrupted by gain errors~$P^{V}$ (red diamonds), calibrated residuals~$P^{R}$ (orange squares), calibrated residuals for a simulation that only contains foregrounds and thermal noise (cyan squares), and a simulation with only thermal noise (magenta circles). Before cross multiplying time samples, the visibilities were coherently averaged over all integrations. {\bf Left Column:} Spherically averaging over all Fourier modes, including those inside the wedge. {\bf Right Column:} Averaging over all Fourier modes outside of the wedge with a 100\,ns buffer. We plot 95\% confidence intervals for the power spectrum of 21\,cm fluctuations~$P^S$ with sample variance as grey regions and denote negative points with open circles. {\bf Bottom Row:} We zoom in on the 21\,cm fluctuations and calibration residuals. Error bars denote 95\% confidence regions. We do not apply any corrective normalization to the power spectrum and only multiply by constant normalization factors constant in $k$. When we average over all modes, including those inside of the wedge, we see that our calibrated residual is biased below $k\lesssim 0.25 h$Mpc$^{-1}$ because of degeneracies between wedge 21\,cm modes and foreground estimates.}\label{fig:SPHERICAL}
\end{figure*}

In Fig.~\ref{fig:SPHERICAL}, we show spherically averaged power spectra for $P^V$, $P^{S+F+N}$ and $P^{S}$ averaging over all sampled Fourier bins. We compare with $P^{R}$ spherically averaged over all Fourier bins lying outside of the wedge plus a 100\,ns buffer. We find that $P^R$ is in broad agreement with with $P^S$ beyond $k \gtrsim 0.2 h^{-1}$Mpc$^{-1}$ when we only spherically average modes outside of the wedge. As shown in \citet{EwallWice:2021}, our residual does significantly better then a tapered estimator (approximated by the black points) which is contaminated by Blackman-Harris sidelobes. When we include modes inside of the wedge in our spherical average, we encounter negative bias out to $\approx 0.3h$Mpc$^{-1}$. This is because we subtracted foreground estimates inside of the wedge which are completely degenerate with 21\,cm modes. 

In Fig.~\ref{fig:SPHERICAL}, we also show calibration residuals from simulations with foregrounds and noise but no EoR. Calibration residuals are an order of magnitude lower then the EoR, at a similar level to the noise only simulations, and $\approx 90$\,dB below the level of the foregrounds in the power-spectrum domain. Our minimally-redundant noise residuals show a slight positive bias that is at roughly the same level as the thermal noise at small k values. We hypothesize that this could  be caused by residual gain errors that still leak foregrounds or artifacts in the simulation itself.

\section{Performance of \acro{} on a Redundant Array}\label{sec:REDUNDANT}
\begin{figure*}
    \centering
    \includegraphics[width=\textwidth]{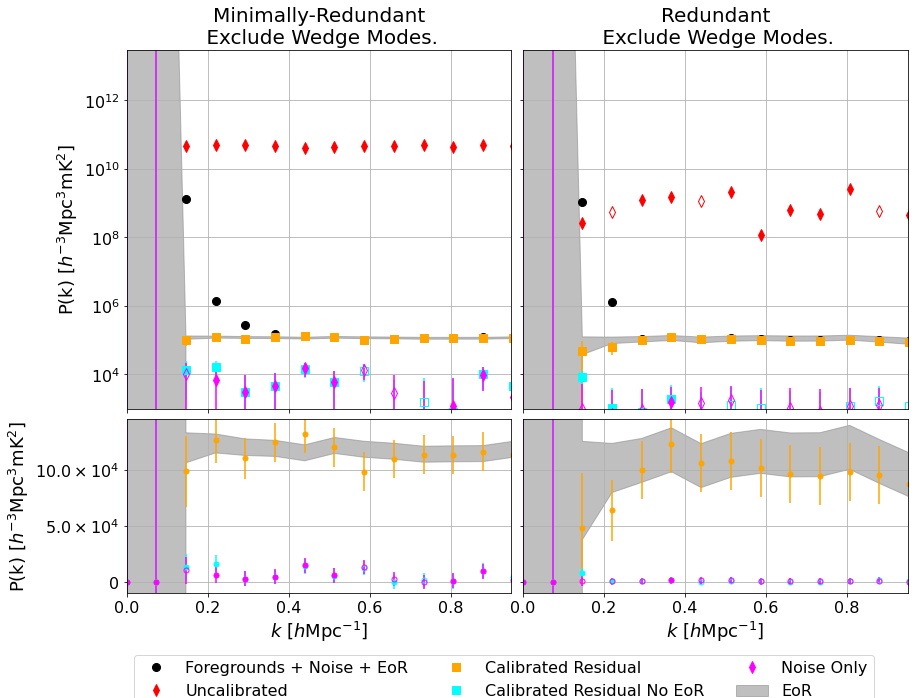}
    \caption{The same as Fig.~\ref{fig:SPHERICAL} except now the left column is our minimally-redundant Array and the right column is our redundant array and we only spherically average Fourier modes outside of the wedge in both the left and right columns. We are able to reduce calibration systematics on our redundant array by 1-2 orders of magnitude by only crossing different redundant baselines to form power spectra. \acro{} is able to recover 21\,cm fluctuations for the redundant case, albeit with higher signal loss (compare the regions below $\lesssim$ 0.3$h$Mpc$^{-1}$ on the left and right plot. This is because fewer independent samples of 21\,cm fluctuations are absorbed into our gain estimates in our redundant array which leads to greater signal loss.}
    \label{fig:SPHERICAL_REDUNDANT}
\end{figure*}

We have demonstrated that on a minimally redundant array \acro{} can remove foreground leakage into the EoR window at dynamic ranges necessary for 21\,cm cosmology, even when our gains contain large amounts of fine frequency structure. In this section, we examine its performance on a highly redundant array (the ``Redundant'' MWA subarray) and compare to the performance on our minimally redundant case. In Fig.~\ref{fig:SPHERICAL_REDUNDANT}, we compare spherical averages of Fourier modes outside of the wedge with a 100\,ns buffer.

We observe that our redundant array suppresses calibration errors by one-to-two orders of magnitude compared to our minimally redundant array. This is because when we average over power spectra formed from pairs of non-identical baselines in each redundant group, we average over independent calibration systematics which average down. We are not able to do the same in our minimally redundant array since each baseline samples a unique sky mode and must be crossed with itself. Because we are able to coherently average over more baselines (rather then incoherently average), the noise-only power spectrum is also significantly reduced for our redundant simulation, relative to our minimally redundant simulation.

Another noteable difference between the redundant and minimally-redundant calibration residuals in Fig.~\ref{fig:SPHERICAL_REDUNDANT} is that the calibration residual in our redundant array has a negative bias at larger k than the minimally-redundant array. This is due to the fact that our gains absorb some 21\,cm signal at all frequency scales which incurs signal loss. Each gain absorbs 21\,cm contributions from the $\sim \Nant$ baselines that each antenna participates in. In our minimally-redundant array, each of these 21\,cm visibilities is statistically independent while in our redundant array, these visibilities are highly correlated within a redundant group, reducing the effective number of independent 21\,cm realizations that are averaged down in each gain solution which increases signal loss on all scales. This effect should be reduced on larger arrays with greater numbers of independent baseline groups involving each gain. It can also be reduced by imposing priors on the gain, such as smoothness, to reduce the number of degrees of freedom we must solve for and their overlap with 21\,cm fluctuations.

\section{The Low Signal-to-Noise regime}\label{sec:LOWSNR}
\begin{figure}
    \centering
    \includegraphics[width=.5\textwidth]{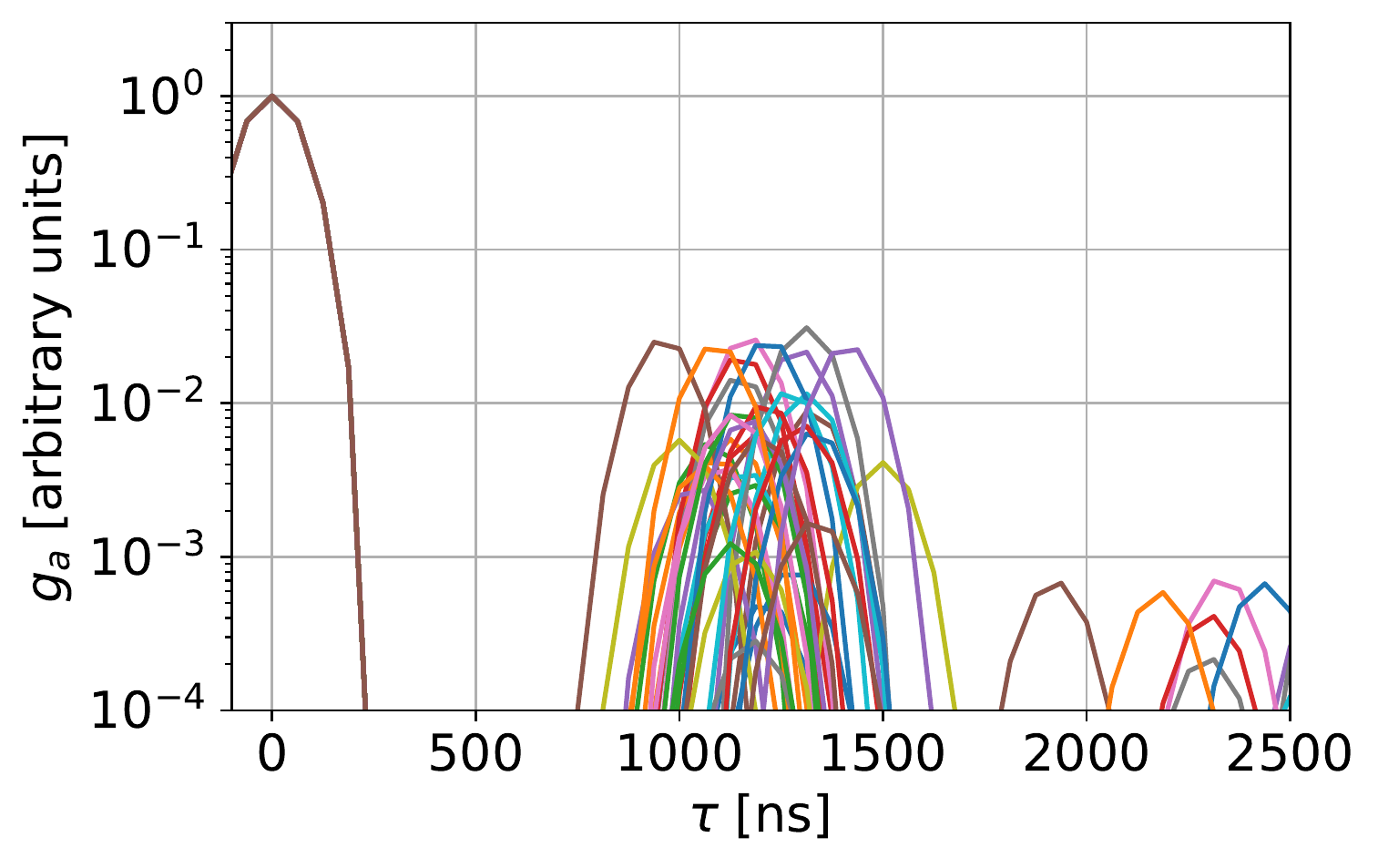}
    \caption{Fourier transforms of simulated gains that we used to test \acro{}'s performance in the low SNR regime. Gains include a single reflection with an amplitude of $\approx 0.01$ and delay of $\approx 1200$\,ns (see equation~\ref{eq:REFGAIN}) and are otherwise flat. Secondary peaks at higher delays are from the second round-trip reflection.}
    \label{fig:DELAYTRANSFORMGAINS}
\end{figure}

\begin{figure}
    \centering
    \includegraphics[width=.5\textwidth]{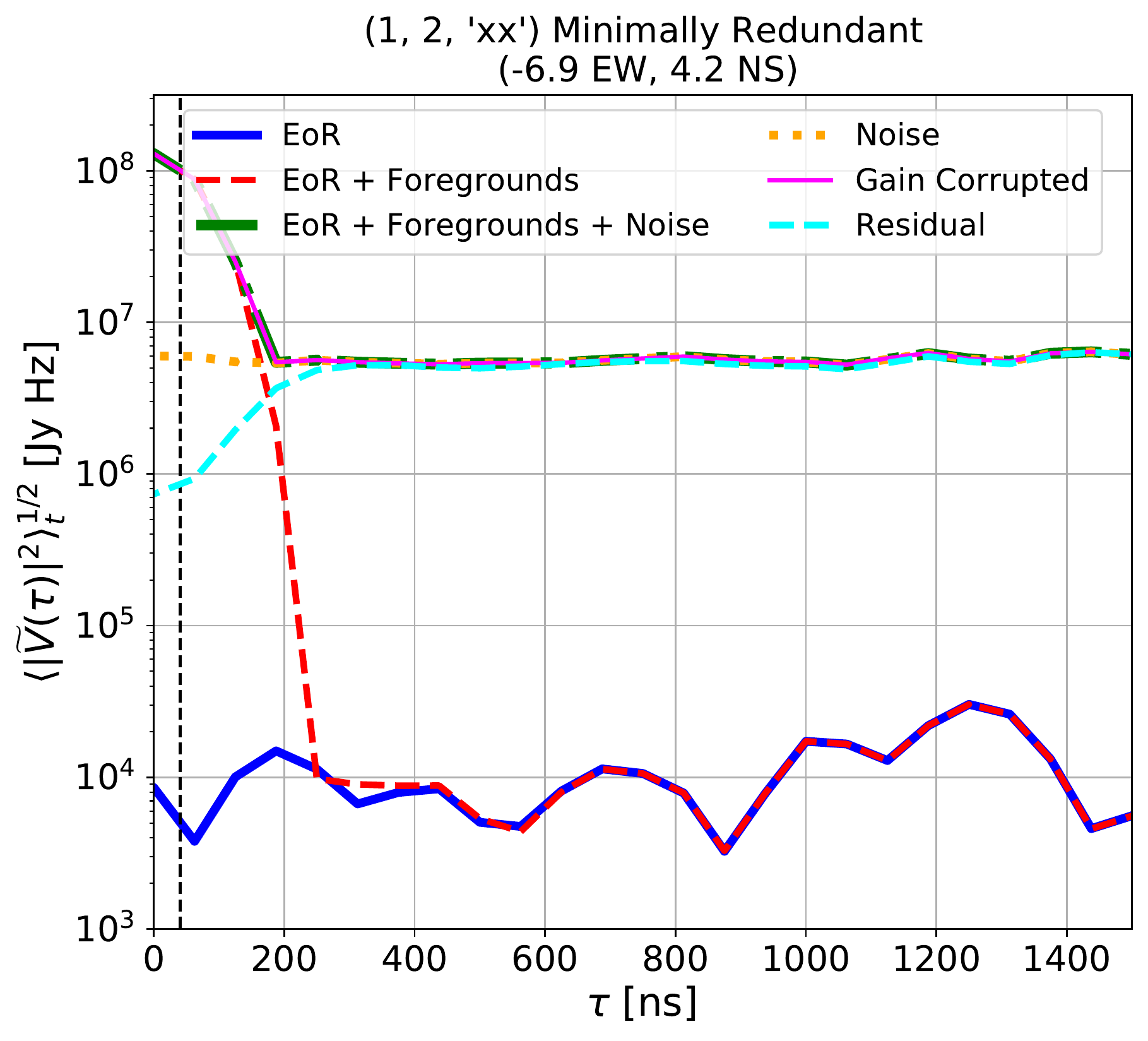}
    \caption{Same as Fig.~\ref{fig:INSPECT_BLS} but now for a single baseline with the level of thermal noise set so that on a single integration, the RMS is equal to one tenth the RMS amplitude of foregrounds. We have also introduced gain errors with distinctive reflection features at the $0.01$ level which are dominated on a single baseline by thermal noise. We clearly observe that the gain corrupted visibilities, residuals, and noise are all at a similar level for a single baseline and integration.}
    \label{fig:INSPECT_BLS_LOW_SNR}
\end{figure}

For the majority of this paper, we have applied \acro{} to simulations with gain-features with much larger amplitudes than the thermal noise. This is because our primary aim is to demonstrate \acro{}'s ability to remove gain features to the level necessary for 21\,cm cosmology. In practice, calibration is usually run on much lower SNR data. In this section, we check whether \acro{} is able to remove gain artifacts to below the level of thermal noise for observations with lower (and more representative) sensitivities. We explore this by observing the power-spectrum of \acro{} residuals for simulations where the true gains are equal to unity plus a small, visually distinctive  sinusoidal ripple that is below the level of thermal noise on a particular integration and baseline. For each antenna $a$, we set 
\begin{equation}\label{eq:REFGAIN}
    g_a = \frac{1}{1 + r_a \text{exp}\left[-I (2 \pi \nu \tau_a + \phi_a)\right]}
\end{equation}
where $r_a$ is an amplitude drawn from an exponential distribution with a mean of $0.01$, $\phi_a$ is a uniformly distributed phase, and $\tau_a$ is drawn from a normal distribution with mean of $1200$\,ns and a standard deviation of $100$\,ns. We show the delay-transoforms of our gains in Fig.~\ref{fig:DELAYTRANSFORMGAINS}. In Fig.~\ref{fig:INSPECT_BLS_LOW_SNR} we show the RMS of a single baseline in delay-space over all integrations. As discussed, the noise level is now roughly an order of magnitude below the foreground level and the gain errors are relatively small as compared to the noise so that on an individual baseline, ``uncalibrated'' visibilities with the corrupting gains are visually difficult to distinguish from the visibilities without the gains. The residuals from \acro{} agree reasonably well with both of these lines. 

\begin{figure}
    \centering
    \includegraphics[width=.5\textwidth]{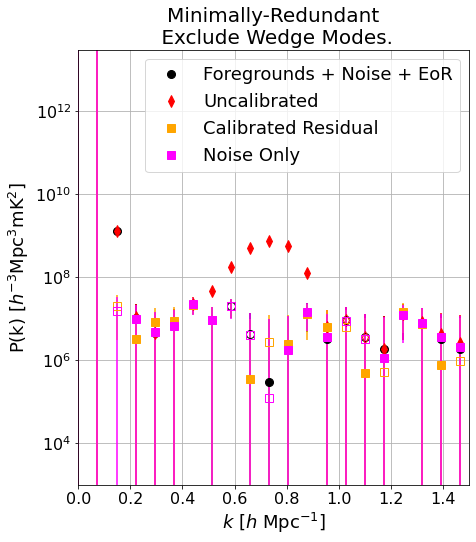}
    \caption{The same as the right-hand panel of Fig.~\ref{fig:SPHERICAL} but now with reflection gains (fig.~\ref{fig:DELAYTRANSFORMGAINS}) and noise that is roughly one-tenth the amplitude of the foregrounds on a single baseline and integration. A reflection features is clearly visible in our uncalibrated visibilities (red points) but it is removed by \acro{} in the residuals (orange squares) which is roughly the same as the residuals for simulations with noise only (purple squares).}
    \label{fig:SPHERICAL_LOW_SNR}
\end{figure}

We are concerned that gain structures below the level of the noise on a single baseline and integration might still be present in the data and raise their heads once we perform averaging over many times and baselines. To determine whether this is the case with \acro{}, we plot the spherically averaged power spectra from all times and baselines in Fig.~\ref{fig:SPHERICAL_LOW_SNR}. With no calibration, reflection artifacts exceed the thermal noise level by a factor of $\approx 10$ in the spherically averaged power spectrum. \acro{} removes the reflection artifacts so that the residual is still consistent with noise only simulations. We determine that in the regime where gain features are below the thermal noise level on a single integration and baseline, \acro{} is still able to mitigate gain-like spectral structures to the noise level of the spherically averaged power spectrum for the entire dataset.

\section{Computational Runtime}\label{sec:COMPUTATION}
\begin{figure}
    \centering
    \includegraphics[width=.5\textwidth]{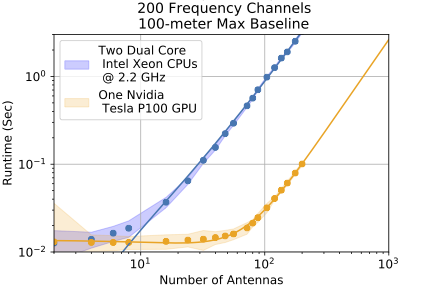}
    \caption{Runtimes for a single \textsc{adamax} gradient update step in \acro{} for a linear array with a maximum baseline length of one-hundred meters (fixed $\Nvec^\text{max}$) and $\Nant$ equally spaced antennas arranged in rows of 8 with 2\,m spacing between the antenna of each row and $\Nfreq=200$ channels over $80$\,MHz and centered at $150$\,MHz. Markers denote distribution medians while filled regions denote 95\% confidence regions. The computational time for a gradient update follows quadratic scaling with $\Nant$, as we expect (lines). \acro{} performs $\gtrsim 30$ times faster on a GPU than on a CPU on arrays with hundreds of antennas (blue regions). Gradient descent typically requires thousands of steps to converge. Thus, on a single GPU equipped computer node \acro{} can provide robust per time-integration bandpass calibration for hundreds of element arrays on the timescales of minutes and for thousands of element arrays on timescales of hours.}
    \label{fig:ANTPOS_RUNTIME}
\end{figure}

\begin{figure}
    \centering
    \includegraphics[width=.5\textwidth]{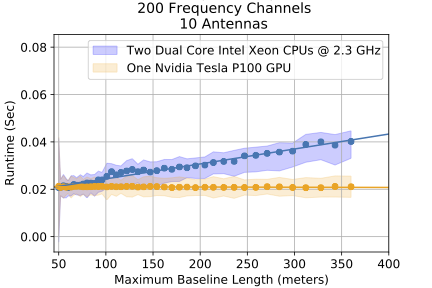}
    \caption{The runtime of a single \textsc{adamax} gradient-descent step for a 10-element linear array of equally spaced antennas with 200 channels between 120 and 180 \,MHz. Shaded regions describe 95\% confidence regions over 1000 realizations on Two dual core Xeon CPUs (blue) and a P100 GPU (orange). We observe linear scaling of the runtime on the CPU. The runtime on the GPU appears constant since we are not yet in the regime where linear scaling is occuring. }
    \label{fig:MAXBL_RUNTIME}
\end{figure}

In this section, we discuss the computational runtime of \acro{}. The computational cost of each gradient update step is dominated by the sum $\sum u_{abi} A_{iab\nu}$ in equation~\ref{eq:LOGLIKELIHOOD} which involves $\Nvec \Nfreq \Nbl$ floating point operations. Memory use is dominated by the $\Nvec \Nfreq \Nbl$ element $\boldsymbol{A}$ matrix. In Fig.~\ref{fig:ANTPOS_RUNTIME}, we show the scaling in runtime for a single \textsc{adamax} gradient update step with the number of antennas in rectangular arrays with rows of eight 2\,m apertures with 2\,m spacing between each antenna in each row and spacings between rows set so that the maximum baseline length is always $100$\,m to keep $\Nvecmax$ constant. Each simulation contains 200 frequency channels between $120-180$\,MHz. We also illustrate the linear scaling of the gradient update computations with maximum baseline length $\propto \Nvecmax$ in Fig.~\ref{fig:MAXBL_RUNTIME} where we hold the number of elements constant (10) in a linear array but change the inter-baseline spacing. We measure runtimes on two different machines, one with Two dual core Intel Xeon CPUs at 2.3\,GHz and a another with identical CPUs and a Tesla P100 GPU. Gradient update steps on the GPU tend to run on the order of $\sim 20-30$\,ms for arrays with $\sim 100$ antennas. We find that given a learning rate of $\gamma = 0.01$, \acro{} requires thousands of gradient descent steps to converge to the precisions required for 21\,cm cosmology. Thus \acro{} can be used to obtain gradient descent solutions for arrays with hundreds of antennas within minutes on a single computer-node outfitted with a modern GPU.


\section{Caveats and Limitations}\label{sec:CAVEATS}
We have demonstrated that \acro{} can remove arbitrary frequency dependent gains at high precision. However, there are several scenarios where its ability to correct systematics will be limited. Firstly, while \acro{} can deal with non-redundancy, it requires that non-redundant systematics only occupy spectral scales within the foreground wedge.  We do not expect \acro{} to be able to correct high-delay systematics that are direction dependent high RM polarization leakage \citep{Moore:2013, Asad:2016} and mutual coupling artifacts that extend significantly beyond the edge of the wedge such as those discussed in \citet{Kern:2019} and \citet{Josaitis:2021}. Thus, telescope designers can relax their specifications on the analog signal chain and its direction independent effects but great care must be still be taken to limit the effects of inter-antenna mutual coupling over large delays. \citet{Josaitis:2021} find that super-horizon emission from mutual coupling primarily contaminates fringe-rates outside of the fringe-rates occupied by the main lobe. Thus main-lobe filtering along the lines of \citet{Parsons:2016} prior to bandpass calibration could still allow for \acro{} to remove fine-frequency bandpass features as long as they are stable on time-scales longer then sky variations.

Another limitation of our technique in it's current form is its reliance on first-order gradient descent. This requires that we make an initial guess of the calibration solution within the convex region containing the local minimum. Thus, we recommend using \acro{} as a final calibration to be performed after a rougher initial calibration 

Finally, because \acro{} models the foregrounds per-baseline, it introduce signal loss within a region close to the wedge plus whatever buffer the analyst has specified. The degree and extent of this signal loss will not only depend on the specific configuration of the interferometer (as we have seen in Fig.~\ref{fig:SPHERICAL}) but will also depend on the bandwidth \citep{EwallWice:2021} and channelization resolution. Care must be taken in any analysis that uses \acro{} to account for this loss either through simulations or propagating the filtered modes into a window function matrix (e.g. \citealt{Kern:2021}).

\section{Conclusion}\label{sec:CONCLUSION}
In this paper, we introduced \acro{}, a fast and spectrally robust calibration algorithm for 21\,cm cosmology. The key innovations leveraged by \acro{} are that it gives up on trying to precisely solve for the highly degenerate gains and foregrounds and instead focuses on solving for and subtracting the products of gains and foregrounds in a way that only relies on the robust assumption that the intrinsic foregrounds are are contained within the wedge region of instrumental Fourier space. We are able to accomplish the computationally expensive modeling problem in a reasonable amount of time for modern array sizes by taking advantage of GPU acceleration and auto-differentiation. By simultaneously modeling gains with per-baseline smooth foreground components, \acro{} is capable of removing arbitrary direction independent bandpass structures to the precision necessary for 21\,cm intensity mapping experiments in the presence of non-redundancy and a highly uncertain foreground sky. We demonstrated \acro{}'s effectiveness using simulations of a redundant and minimally-redundant configuration on the Murchison Widefield Array and find that in both scenarios, it recovers relatively unbiased estimates of the power spectrum within the EoR window. While we assumed artifically high SNR to demonstrate \acro{}'s ability to filter out systematics to the level of 21\,cm fluctuations, we also find that \acro{} removes spectral gain errors that are below the noise level of a single baseline integration and is effective in the low SNR regime encountered in the field. Low-level spectral bandpass structures are one of the primary systematics currently hampering observational efforts, thus \acro{} has the potential to drive significant progress towards a detection.  Because \acro{} models each baseline independently, it does not require redundancy which can be difficult to achieve in the field but as a consequence, it gives up on recovering any 21\,cm signal inside of the foreground wedge since its per-baseline spectral modes are completely degenerate with large LoS scale 21\,cm fluctuations. While \acro{} can correct arbitrary direction independent gain errors, we do not expect it to be capable of correcting direction dependent gain errors with delay structures that extend significantly beyond the edge of the wedge. Thus, array designers should focus their efforts on keeping direction dependent over-the-air mutual coupling effects to large spectral scales while potentially relaxing their requirements on redundancy on large spectral scales and direction independent fine-scale features.  
\section{Acknowledgements}
We thank Jacqueline Hewitt for helpful conversations. AEW agknowledges support from the Berkeley Center for Cosmological Physics. This material is based upon work supported by the National Science Foundation under Grant Nos. 1636646 and 1836019 and institutional support from the HERA collaboration partners. This research is funded by the Gordon and Betty Moore Foundation
through grant GBMF5215 to the Massachusetts Institute of Technology.
\section{Code and Data Products}
The \acro{} library is freely available as a git repository at \url{https://github.com/aewallwi/calamity/tree/main/calamity}. All plotting notebooks and larger simulation files used in this work can be obtained by contacting the authors.
This work made extensive use of the $\textsc{tensorflow}$\footnote{\url{https://www.tensorflow.org/}} \citep{Tensorflow:2015}, $\textsc{numpy}$\footnote{\url{https://numpy.org/}} \citep{Numpy:2020}, $\textsc{scipy}$\footnote{\url{https://www.scipy.org/}} \citep{Scipy:2020}, and $\textsc{matplotlib}$\footnote{\url{https://matplotlib.org/}} \citep{Matplotlib:2007} libraries. 
Our visibility simulations were performed using $\textsc{pyuvsim}$\footnote{\url{https://github.com/RadioAstronomySoftwareGroup/pyuvsim}} and their products were stored and manipulated using the data formats and methods from the $\textsc{pyuvdata}$\footnote{\url{https://github.com/RadioAstronomySoftwareGroup/pyuvdata}} library \citep{Pyuvdata:2017}. Power-spectrum calculations were performed using the $\textsc{hera\_pspec}$ library\footnote{\url{https://github.com/HERA-Team/hera_pspec}}.
\bibliography{calamity}

\begin{thebibliography}{}
\expandafter\ifx\csname natexlab\endcsname\relax\def\natexlab#1{#1}\fi
\providecommand{\url}[1]{\href{#1}{#1}}
\providecommand{\dodoi}[1]{doi:~\href{http://doi.org/#1}{\nolinkurl{#1}}}
\providecommand{\doeprint}[1]{\href{http://ascl.net/#1}{\nolinkurl{http://ascl.net/#1}}}
\providecommand{\doarXiv}[1]{\href{https://arxiv.org/abs/#1}{\nolinkurl{https://arxiv.org/abs/#1}}}

\bibitem[{Abadi {et~al.}(2015)Abadi, Agarwal, Barham, Brevdo, Chen, Citro,
  Corrado, Davis, Dean, Devin, Ghemawat, Goodfellow, Harp, Irving, Isard, Jia,
  Jozefowicz, Kaiser, Kudlur, Levenberg, Man\'{e}, Monga, Moore, Murray, Olah,
  Schuster, Shlens, Steiner, Sutskever, Talwar, Tucker, Vanhoucke, Vasudevan,
  Vi\'{e}gas, Vinyals, Warden, Wattenberg, Wicke, Yu, \&
  Zheng}]{Tensorflow:2015}
Abadi, M., Agarwal, A., Barham, P., {et~al.} 2015, {TensorFlow}: Large-Scale
  Machine Learning on Heterogeneous Systems.
\newblock \url{https://www.tensorflow.org/}

\bibitem[{{Ali} {et~al.}(2015){Ali}, {Parsons}, {Zheng}, {Pober}, {Liu},
  {Aguirre}, {Bradley}, {Bernardi}, {Carilli}, {Cheng}, {DeBoer}, {Dexter},
  {Grobbelaar}, {Horrell}, {Jacobs}, {Klima}, {MacMahon}, {Maree}, {Moore},
  {Razavi}, {Stefan}, {Walbrugh}, \& {Walker}}]{Ali:2015}
{Ali}, Z.~S., {Parsons}, A.~R., {Zheng}, H., {et~al.} 2015, \apj, 809, 61,
  \dodoi{10.1088/0004-637X/809/1/61}

\bibitem[{{Anderson} {et~al.}(2018){Anderson}, {Luciw}, {Li}, {Kuo}, {Yadav},
  {Masui}, {Chang}, {Chen}, {Oppermann}, {Liao}, {Pen}, {Price},
  {Staveley-Smith}, {Switzer}, {Timbie}, \& {Wolz}}]{Anderson:2018}
{Anderson}, C.~J., {Luciw}, N.~J., {Li}, Y.~C., {et~al.} 2018, \mnras, 476,
  3382, \dodoi{10.1093/mnras/sty346}

\bibitem[{{Asad} {et~al.}(2018){Asad}, {Koopmans}, {Jeli{\'c}}, {de Bruyn},
  {Pandey}, \& {Gehlot}}]{Asad:2018}
{Asad}, K.~M.~B., {Koopmans}, L.~V.~E., {Jeli{\'c}}, V., {et~al.} 2018, \mnras,
  476, 3051, \dodoi{10.1093/mnras/sty258}

\bibitem[{{Asad} {et~al.}(2016){Asad}, {Koopmans}, {Jeli{\'c}}, {Ghosh},
  {Abdalla}, {Brentjens}, {de Bruyn}, {Ciardi}, {Gehlot}, {Iliev}, {Mevius},
  {Pandey}, {Yatawatta}, \& {Zaroubi}}]{Asad:2016}
---. 2016, \mnras, 462, 4482, \dodoi{10.1093/mnras/stw1863}

\bibitem[{{Baars} {et~al.}(1965){Baars}, {Mezger}, \& {Wendker}}]{Baars:1965}
{Baars}, J.~W.~M., {Mezger}, P.~G., \& {Wendker}, H. 1965, \apj, 142, 122,
  \dodoi{10.1086/148267}

\bibitem[{{Bandura} {et~al.}(2014){Bandura}, {Addison}, {Amiri}, {Bond},
  {Campbell-Wilson}, {Connor}, {Cliche}, {Davis}, {Deng}, {Denman}, {Dobbs},
  {Fandino}, {Gibbs}, {Gilbert}, {Halpern}, {Hanna}, {Hincks}, {Hinshaw},
  {H{\"o}fer}, {Klages}, {Landecker}, {Masui}, {Mena Parra}, {Newburgh}, {Pen},
  {Peterson}, {Recnik}, {Shaw}, {Sigurdson}, {Sitwell}, {Smecher}, {Smegal},
  {Vanderlinde}, \& {Wiebe}}]{Bandura:2014}
{Bandura}, K., {Addison}, G.~E., {Amiri}, M., {et~al.} 2014, in Society of
  Photo-Optical Instrumentation Engineers (SPIE) Conference Series, Vol. 9145,
  Ground-based and Airborne Telescopes V, ed. L.~M. {Stepp}, R.~{Gilmozzi}, \&
  H.~J. {Hall}, 914522, \dodoi{10.1117/12.2054950}

\bibitem[{{Barry} {et~al.}(2019){Barry}, {Beardsley}, {Byrne}, {Hazelton},
  {Morales}, {Pober}, \& {Sullivan}}]{Barry:2019b}
{Barry}, N., {Beardsley}, A.~P., {Byrne}, R., {et~al.} 2019, \pasa, 36, e026,
  \dodoi{10.1017/pasa.2019.21}

\bibitem[{{Barry} {et~al.}(2016){Barry}, {Hazelton}, {Sullivan}, {Morales}, \&
  {Pober}}]{Barry:2016}
{Barry}, N., {Hazelton}, B., {Sullivan}, I., {Morales}, M.~F., \& {Pober},
  J.~C. 2016, \mnras, 461, 3135, \dodoi{10.1093/mnras/stw1380}

\bibitem[{{Beardsley} {et~al.}(2016){Beardsley}, {Hazelton}, {Sullivan},
  {Carroll}, {Barry}, {Rahimi}, {Pindor}, {Trott}, {Line}, {Jacobs}, {Morales},
  {Pober}, {Bernardi}, {Bowman}, {Busch}, {Briggs}, {Cappallo}, {Corey}, {de
  Oliveira-Costa}, {Dillon}, {Emrich}, {Ewall-Wice}, {Feng}, {Gaensler},
  {Goeke}, {Greenhill}, {Hewitt}, {Hurley-Walker}, {Johnston-Hollitt},
  {Kaplan}, {Kasper}, {Kim}, {Kratzenberg}, {Lenc}, {Loeb}, {Lonsdale},
  {Lynch}, {McKinley}, {McWhirter}, {Mitchell}, {Morgan}, {Neben},
  {Thyagarajan}, {Oberoi}, {Offringa}, {Ord}, {Paul}, {Prabu}, {Procopio},
  {Riding}, {Rogers}, {Roshi}, {Udaya Shankar}, {Sethi}, {Srivani},
  {Subrahmanyan}, {Tegmark}, {Tingay}, {Waterson}, {Wayth}, {Webster},
  {Whitney}, {Williams}, {Williams}, {Wu}, \& {Wyithe}}]{Beardsley:2016}
{Beardsley}, A.~P., {Hazelton}, B.~J., {Sullivan}, I.~S., {et~al.} 2016, \apj,
  833, 102, \dodoi{10.3847/1538-4357/833/1/102}

\bibitem[{{Berger} {et~al.}(2016){Berger}, {Newburgh}, {Amiri}, {Bandura},
  {Cliche}, {Connor}, {Deng}, {Denman}, {Dobbs}, {Fandino}, {Gilbert}, {Good},
  {Halpern}, {Hanna}, {Hincks}, {Hinshaw}, {H{\"o}fer}, {Johnson}, {Landecker},
  {Masui}, {Mena Parra}, {Oppermann}, {Pen}, {Peterson}, {Recnik}, {Robishaw},
  {Shaw}, {Siegel}, {Sigurdson}, {Smith}, {Storer}, {Tretyakov}, {Van Gassen},
  {Vanderlinde}, \& {Wiebe}}]{Berger:2016}
{Berger}, P., {Newburgh}, L.~B., {Amiri}, M., {et~al.} 2016, in Society of
  Photo-Optical Instrumentation Engineers (SPIE) Conference Series, Vol. 9906,
  Ground-based and Airborne Telescopes VI, ed. H.~J. {Hall}, R.~{Gilmozzi}, \&
  H.~K. {Marshall}, 99060D, \dodoi{10.1117/12.2233782}

\bibitem[{{Byrne} {et~al.}(2021{\natexlab{a}}){Byrne}, {Morales}, {Hazelton},
  {Sullivan}, {Barry}, {Lynch}, {Line}, \& {Jacobs}}]{Byrne:2021b}
{Byrne}, R., {Morales}, M.~F., {Hazelton}, B., {et~al.} 2021{\natexlab{a}},
  arXiv e-prints, arXiv:2107.11487.
\newblock \doarXiv{2107.11487}

\bibitem[{{Byrne} {et~al.}(2021{\natexlab{b}}){Byrne}, {Morales}, {Hazelton},
  \& {Wilensky}}]{Byrne:2021}
{Byrne}, R., {Morales}, M.~F., {Hazelton}, B.~J., \& {Wilensky}, M.
  2021{\natexlab{b}}, \mnras, 503, 2457, \dodoi{10.1093/mnras/stab647}

\bibitem[{{Byrne} {et~al.}(2019){Byrne}, {Morales}, {Hazelton}, {Li}, {Barry},
  {Beardsley}, {Joseph}, {Pober}, {Sullivan}, \& {Trott}}]{Byrne:2019}
{Byrne}, R., {Morales}, M.~F., {Hazelton}, B., {et~al.} 2019, \apj, 875, 70,
  \dodoi{10.3847/1538-4357/ab107d}

\bibitem[{{Chapman} {et~al.}(2013){Chapman}, {Abdalla}, {Bobin}, {Starck},
  {Harker}, {Jeli{\'c}}, {Labropoulos}, {Zaroubi}, {Brentjens}, {de Bruyn}, \&
  {Koopmans}}]{Chapman:2013}
{Chapman}, E., {Abdalla}, F.~B., {Bobin}, J., {et~al.} 2013, \mnras, 429, 165,
  \dodoi{10.1093/mnras/sts333}

\bibitem[{{Chen}(2015)}]{Chen:2015}
{Chen}, X. 2015, in IAU General Assembly, Vol.~29, 2252187

\bibitem[{{Choudhuri} {et~al.}(2021){Choudhuri}, {Bull}, \&
  {Garsden}}]{Choudhuri:2021}
{Choudhuri}, S., {Bull}, P., \& {Garsden}, H. 2021, \mnras,
  \dodoi{10.1093/mnras/stab1795}

\bibitem[{{Colegate} {et~al.}(2015){Colegate}, {Sutinjo}, {Hall}, {Padhi},
  {Wayth}, {Bij de Vaate}, {Crosse}, {Emrich}, {Faulkner}, {Hurley-Walker}, {de
  Lera Acedo}, {Juswardy}, {Razavi-Ghods}, {Tingay}, \&
  {Williams}}]{Colgate:2015}
{Colegate}, T.~M., {Sutinjo}, A.~T., {Hall}, P.~J., {et~al.} 2015, arXiv
  e-prints, arXiv:1501.05017.
\newblock \doarXiv{1501.05017}

\bibitem[{{Datta} {et~al.}(2010){Datta}, {Bowman}, \& {Carilli}}]{Datta:2010}
{Datta}, A., {Bowman}, J.~D., \& {Carilli}, C.~L. 2010, \apj, 724, 526,
  \dodoi{10.1088/0004-637X/724/1/526}

\bibitem[{{de Lera Acedo} {et~al.}(2018){de Lera Acedo}, {Bolli}, {Paonessa},
  {Virone}, {Colin-Beltran}, {Razavi-Ghods}, {Aicardi}, {Lingua}, {Maschio},
  {Monari}, {Naldi}, {Piras}, \& {Pupillo}}]{deLeraAcedo:2018}
{de Lera Acedo}, E., {Bolli}, P., {Paonessa}, F., {et~al.} 2018, Experimental
  Astronomy, 45, 1, \dodoi{10.1007/s10686-017-9566-x}

\bibitem[{{DeBoer} {et~al.}(2017){DeBoer}, {Parsons}, {Aguirre}, {Alexander},
  {Ali}, {Beardsley}, {Bernardi}, {Bowman}, {Bradley}, {Carilli}, {Cheng}, {de
  Lera Acedo}, {Dillon}, {Ewall-Wice}, {Fadana}, {Fagnoni}, {Fritz},
  {Furlanetto}, {Glendenning}, {Greig}, {Grobbelaar}, {Hazelton}, {Hewitt},
  {Hickish}, {Jacobs}, {Julius}, {Kariseb}, {Kohn}, {Lekalake}, {Liu}, {Loots},
  {MacMahon}, {Malan}, {Malgas}, {Maree}, {Martinot}, {Mathison}, {Matsetela},
  {Mesinger}, {Morales}, {Neben}, {Patra}, {Pieterse}, {Pober}, {Razavi-Ghods},
  {Ringuette}, {Robnett}, {Rosie}, {Sell}, {Smith}, {Syce}, {Tegmark},
  {Thyagarajan}, {Williams}, \& {Zheng}}]{DeBoer:2017}
{DeBoer}, D.~R., {Parsons}, A.~R., {Aguirre}, J.~E., {et~al.} 2017, \pasp, 129,
  045001, \dodoi{10.1088/1538-3873/129/974/045001}

\bibitem[{{Di Matteo} {et~al.}(2004){Di Matteo}, {Ciardi}, \&
  {Miniati}}]{DiMatteo:2004}
{Di Matteo}, T., {Ciardi}, B., \& {Miniati}, F. 2004, \mnras, 355, 1053,
  \dodoi{10.1111/j.1365-2966.2004.08443.x}

\bibitem[{{Dillon} {et~al.}(2014){Dillon}, {Liu}, {Williams}, {Hewitt},
  {Tegmark}, {Morgan}, {Levine}, {Morales}, {Tingay}, {Bernardi}, {Bowman},
  {Briggs}, {Cappallo}, {Emrich}, {Mitchell}, {Oberoi}, {Prabu}, {Wayth}, \&
  {Webster}}]{Dillon:2014}
{Dillon}, J.~S., {Liu}, A., {Williams}, C.~L., {et~al.} 2014, \prd, 89, 023002,
  \dodoi{10.1103/PhysRevD.89.023002}

\bibitem[{{Dillon} {et~al.}(2018){Dillon}, {Kohn}, {Parsons}, {Aguirre}, {Ali},
  {Bernardi}, {Kern}, {Li}, {Liu}, {Nunhokee}, \& {Pober}}]{Dillon:2018}
{Dillon}, J.~S., {Kohn}, S.~A., {Parsons}, A.~R., {et~al.} 2018, \mnras, 477,
  5670, \dodoi{10.1093/mnras/sty1060}

\bibitem[{{Dillon} {et~al.}(2020){Dillon}, {Lee}, {Ali}, {Parsons}, {Orosz},
  {Nunhokee}, {La Plante}, {Beardsley}, {Kern}, {Abdurashidova}, {Aguirre},
  {Alexander}, {Balfour}, {Bernardi}, {Billings}, {Bowman}, {Bradley}, {Bull},
  {Burba}, {Carey}, {Carilli}, {Cheng}, {DeBoer}, {Dexter}, {de Lera Acedo},
  {Ely}, {Ewall-Wice}, {Fagnoni}, {Fritz}, {Furlanetto}, {Gale-Sides},
  {Glendenning}, {Gorthi}, {Greig}, {Grobbelaar}, {Halday}, {Hazelton},
  {Hewitt}, {Hickish}, {Jacobs}, {Julius}, {Kerrigan}, {Kittiwisit}, {Kohn},
  {Kolopanis}, {Lanman}, {Lekalake}, {Lewis}, {Liu}, {Ma}, {MacMahon}, {Malan},
  {Malgas}, {Maree}, {Martinot}, {Matsetela}, {Mesinger}, {Molewa}, {Morales},
  {Mosiane}, {Murray}, {Neben}, {Nikolic}, {Pascua}, {Patra}, {Pieterse},
  {Pober}, {Razavi-Ghods}, {Ringuette}, {Robnett}, {Rosie}, {Santos}, {Sims},
  {Smith}, {Syce}, {Tegmark}, {Thyagarajan}, {Williams}, \&
  {Zheng}}]{Dillon:2020}
{Dillon}, J.~S., {Lee}, M., {Ali}, Z.~S., {et~al.} 2020, \mnras, 499, 5840,
  \dodoi{10.1093/mnras/staa3001}

\bibitem[{{Eastwood} {et~al.}(2019){Eastwood}, {Anderson}, {Monroe},
  {Hallinan}, {Catha}, {Dowell}, {Garsden}, {Greenhill}, {Hicks}, {Kocz},
  {Price}, {Schinzel}, {Vedantham}, \& {Wang}}]{Eastwood:2019}
{Eastwood}, M.~W., {Anderson}, M.~M., {Monroe}, R.~M., {et~al.} 2019, \aj, 158,
  84, \dodoi{10.3847/1538-3881/ab2629}

\bibitem[{{Ewall-Wice} {et~al.}(2017){Ewall-Wice}, {Dillon}, {Liu}, \&
  {Hewitt}}]{EwallWice:2017}
{Ewall-Wice}, A., {Dillon}, J.~S., {Liu}, A., \& {Hewitt}, J. 2017, \mnras,
  470, 1849, \dodoi{10.1093/mnras/stx1221}

\bibitem[{{Ewall-Wice} {et~al.}(2016){Ewall-Wice}, {Bradley}, {Deboer},
  {Hewitt}, {Parsons}, {Aguirre}, {Ali}, {Bowman}, {Cheng}, {Neben}, {Patra},
  {Thyagarajan}, {Venter}, {de Lera Acedo}, {Dillon}, {Dickenson}, {Doolittle},
  {Egan}, {Hedrick}, {Klima}, {Kohn}, {Schaffner}, {Shelton}, {Saliwanchik},
  {Taylor}, {Taylor}, {Tegmark}, \& {Wirt}}]{EwallWice:2016b}
{Ewall-Wice}, A., {Bradley}, R., {Deboer}, D., {et~al.} 2016, \apj, 831, 196,
  \dodoi{10.3847/0004-637X/831/2/196}

\bibitem[{{Ewall-Wice} {et~al.}(2021){Ewall-Wice}, {Kern}, {Dillon}, {Liu},
  {Parsons}, {Singh}, {Lanman}, {Plante}, {Fagnoni}, {Acedo}, {DeBoer},
  {Nunhokee}, {Bull}, {Chang}, {Lazio}, {Aguirre}, \&
  {Weinberg}}]{EwallWice:2021}
{Ewall-Wice}, A., {Kern}, N., {Dillon}, J.~S., {et~al.} 2021, \mnras, 500,
  5195, \dodoi{10.1093/mnras/staa3293}

\bibitem[{{Fagnoni} {et~al.}(2021){Fagnoni}, {de Lera Acedo}, {DeBoer},
  {Abdurashidova}, {Aguirre}, {Alexander}, {Ali}, {Balfour}, {Beardsley},
  {Bernardi}, {Billings}, {Bowman}, {Bradley}, {Bull}, {Burba}, {Carilli},
  {Cheng}, {Dexter}, {Dillon}, {Ewall-Wice}, {Fritz}, {Furlanetto},
  {Gale-Sides}, {Glendenning}, {Gorthi}, {Greig}, {Grobbelaar}, {Halday},
  {Hazelton}, {Hewitt}, {Hickish}, {Jacobs}, {Josaitis}, {Julius}, {Kern},
  {Kerrigan}, {Kim}, {Kittiwisit}, {Kohn}, {Kolopanis}, {Lanman}, {Plante},
  {Lekalake}, {Liu}, {MacMahon}, {Malan}, {Malgas}, {Maree}, {Martinot},
  {Matsetela}, {Mena Parra}, {Mesinger}, {Molewa}, {Morales}, {Mosiane},
  {Neben}, {Nikolic}, {Parsons}, {Patra}, {Pieterse}, {Pober}, {Razavi-Ghods},
  {Robnett}, {Rosie}, {Sims}, {Smith}, {Syce}, {Thyagarajan}, {Williams}, \&
  {Zheng}}]{Fagnoni:2021}
{Fagnoni}, N., {de Lera Acedo}, E., {DeBoer}, D.~R., {et~al.} 2021, \mnras,
  500, 1232, \dodoi{10.1093/mnras/staa3268}

\bibitem[{{Gehlot} {et~al.}(2018){Gehlot}, {Koopmans}, {de Bruyn}, {Zaroubi},
  {Brentjens}, {Asad}, {Hatef}, {Jeli{\'c}}, {Mevius}, {Offringa}, {Pandey}, \&
  {Yatawatta}}]{Gehlot:2018}
{Gehlot}, B.~K., {Koopmans}, L.~V.~E., {de Bruyn}, A.~G., {et~al.} 2018,
  \mnras, 478, 1484, \dodoi{10.1093/mnras/sty1095}

\bibitem[{{Gehlot} {et~al.}(2019){Gehlot}, {Mertens}, {Koopmans}, {Brentjens},
  {Zaroubi}, {Ciardi}, {Ghosh}, {Hatef}, {Iliev}, {Jeli{\'c}}, {}, {Kooistra},
  {Krause}, {Mellema}, {Mevius}, {Mitra}, {Offringa}, {Pandey}, {Sardarabadi},
  {Schaye}, {Silva}, {Vedantham}, \& {Yatawatta}}]{Gehlot:2019}
{Gehlot}, B.~K., {Mertens}, F.~G., {Koopmans}, L.~V.~E., {et~al.} 2019, \mnras,
  488, 4271, \dodoi{10.1093/mnras/stz1937}

\bibitem[{{Gu} {et~al.}(2013){Gu}, {Xu}, {Wang}, {An}, \& {Chen}}]{Gu:2013}
{Gu}, J., {Xu}, H., {Wang}, J., {An}, T., \& {Chen}, W. 2013, \apj, 773, 38,
  \dodoi{10.1088/0004-637X/773/1/38}

\bibitem[{{Gueuning} {et~al.}(2020){Gueuning}, {Brown}, {Craeye}, \& {de Lera
  Acedo}}]{Gueuning:2020}
{Gueuning}, Q., {Brown}, A., {Craeye}, C., \& {de Lera Acedo}, E. 2020, arXiv
  e-prints, arXiv:2005.11712.
\newblock \doarXiv{2005.11712}

\bibitem[{Harris {et~al.}(2020)Harris, Millman, van~der Walt, Gommers,
  Virtanen, Cournapeau, Wieser, Taylor, Berg, Smith, Kern, Picus, Hoyer, van
  Kerkwijk, Brett, Haldane, del R{\'{i}}o, Wiebe, Peterson,
  G{\'{e}}rard-Marchant, Sheppard, Reddy, Weckesser, Abbasi, Gohlke, \&
  Oliphant}]{Numpy:2020}
Harris, C.~R., Millman, K.~J., van~der Walt, S.~J., {et~al.} 2020, Nature, 585,
  357, \dodoi{10.1038/s41586-020-2649-2}

\bibitem[{{Hazelton} {et~al.}(2017){Hazelton}, {Jacobs}, {Pober}, \&
  {Beardsley}}]{Pyuvdata:2017}
{Hazelton}, B.~J., {Jacobs}, D.~C., {Pober}, J.~C., \& {Beardsley}, A.~P. 2017,
  The Journal of Open Source Software, 2, 140, \dodoi{10.21105/joss.00140}

\bibitem[{Hunter(2007)}]{Matplotlib:2007}
Hunter, J.~D. 2007, Computing in Science \& Engineering, 9, 90,
  \dodoi{10.1109/MCSE.2007.55}

\bibitem[{Hurley-Walker {et~al.}(2016)Hurley-Walker, Callingham, Hancock,
  Franzen, Hindson, Kapińska, Morgan, Offringa, Wayth, Wu, Zheng, Murphy,
  Bell, Dwarakanath, For, Gaensler, Johnston-Hollitt, Lenc, Procopio,
  Staveley-Smith, Ekers, Bowman, Briggs, Cappallo, Deshpande, Greenhill,
  Hazelton, Kaplan, Lonsdale, McWhirter, Mitchell, Morales, Morgan, Oberoi,
  Ord, Prabu, Shankar, Srivani, Subrahmanyan, Tingay, Webster, Williams, \&
  Williams}]{Hurley-Walker:2016}
Hurley-Walker, N., Callingham, J.~R., Hancock, P.~J., {et~al.} 2016, \mnras,
  464, 1146, \dodoi{10.1093/mnras/stw2337}

\bibitem[{{Intema} {et~al.}(2009){Intema}, {van der Tol}, {Cotton}, {Cohen},
  {van Bemmel}, \& {R{\"o}ttgering}}]{Intema:2009}
{Intema}, H.~T., {van der Tol}, S., {Cotton}, W.~D., {et~al.} 2009, \aap, 501,
  1185, \dodoi{10.1051/0004-6361/200811094}

\bibitem[{{Jacobs} {et~al.}(2013){Jacobs}, {Parsons}, {Aguirre}, {Ali},
  {Bowman}, {Bradley}, {Carilli}, {DeBoer}, {Dexter}, {Gugliucci}, {Klima},
  {MacMahon}, {Manley}, {Moore}, {Pober}, {Stefan}, \&
  {Walbrugh}}]{Jacobs:2013}
{Jacobs}, D.~C., {Parsons}, A.~R., {Aguirre}, J.~E., {et~al.} 2013, \apj, 776,
  108, \dodoi{10.1088/0004-637X/776/2/108}

\bibitem[{{Jacobs} {et~al.}(2017){Jacobs}, {Burba}, {Bowman}, {Neben},
  {Stinnett}, {Turner}, {Johnson}, {Busch}, {Allison}, {Leatham}, {Serrano
  Rodriguez}, {Denney}, \& {Nelson}}]{Jacobs:2017}
{Jacobs}, D.~C., {Burba}, J., {Bowman}, J.~D., {et~al.} 2017, \pasp, 129,
  035002, \dodoi{10.1088/1538-3873/aa56b9}

\bibitem[{{Jordan} {et~al.}(2017){Jordan}, {Murray}, {Trott}, {Wayth},
  {Mitchell}, {Rahimi}, {Pindor}, {Procopio}, \& {Morgan}}]{Jordan:2017}
{Jordan}, C.~H., {Murray}, S., {Trott}, C.~M., {et~al.} 2017, \mnras, 471,
  3974, \dodoi{10.1093/mnras/stx1797}

\bibitem[{{Josaitis} {et~al.}(in preparation.)}]{Josaitis:2021}
{Josaitis}, A.~C., {et~al.} in preparation.

\bibitem[{{Joseph} {et~al.}(2018){Joseph}, {Trott}, \& {Wayth}}]{Joseph:2018}
{Joseph}, R.~C., {Trott}, C.~M., \& {Wayth}, R.~B. 2018, \aj, 156, 285,
  \dodoi{10.3847/1538-3881/aaec0b}

\bibitem[{{Joseph} {et~al.}(2020){Joseph}, {Trott}, {Wayth}, \&
  {Nasirudin}}]{Joseph:2020}
{Joseph}, R.~C., {Trott}, C.~M., {Wayth}, R.~B., \& {Nasirudin}, A. 2020,
  \mnras, 492, 2017, \dodoi{10.1093/mnras/stz3375}

\bibitem[{{Kazemi} \& {Yatawatta}(2013)}]{Kazemi:2013}
{Kazemi}, S., \& {Yatawatta}, S. 2013, \mnras, 435, 597,
  \dodoi{10.1093/mnras/stt1347}

\bibitem[{{Kazemi} {et~al.}(2011){Kazemi}, {Yatawatta}, {Zaroubi},
  {Lampropoulos}, {de Bruyn}, {Koopmans}, \& {Noordam}}]{Kazemi:2011}
{Kazemi}, S., {Yatawatta}, S., {Zaroubi}, S., {et~al.} 2011, \mnras, 414, 1656,
  \dodoi{10.1111/j.1365-2966.2011.18506.x}

\bibitem[{{Kenyon} {et~al.}(2018){Kenyon}, {Smirnov}, {Grobler}, \&
  {Perkins}}]{Kenyon:2018}
{Kenyon}, J.~S., {Smirnov}, O.~M., {Grobler}, T.~L., \& {Perkins}, S.~J. 2018,
  \mnras, 478, 2399, \dodoi{10.1093/mnras/sty1221}

\bibitem[{{Kern} \& {Liu}(2021)}]{Kern:2021}
{Kern}, N.~S., \& {Liu}, A. 2021, \mnras, 501, 1463,
  \dodoi{10.1093/mnras/staa3736}

\bibitem[{{Kern} {et~al.}(2019){Kern}, {Parsons}, {Dillon}, {Lanman},
  {Fagnoni}, \& {de Lera Acedo}}]{Kern:2019}
{Kern}, N.~S., {Parsons}, A.~R., {Dillon}, J.~S., {et~al.} 2019, \apj, 884,
  105, \dodoi{10.3847/1538-4357/ab3e73}

\bibitem[{{Kern} {et~al.}(2020{\natexlab{a}}){Kern}, {Parsons}, {Dillon},
  {Lanman}, {Liu}, {Bull}, {Ewall-Wice}, {Abdurashidova}, {Aguirre},
  {Alexander}, {Ali}, {Balfour}, {Beardsley}, {Bernardi}, {Bowman}, {Bradley},
  {Burba}, {Carilli}, {Cheng}, {DeBoer}, {Dexter}, {de Lera Acedo}, {Fagnoni},
  {Fritz}, {Furlanetto}, {Glendenning}, {Gorthi}, {Greig}, {Grobbelaar},
  {Halday}, {Hazelton}, {Hewitt}, {Hickish}, {Jacobs}, {Julius}, {Kerrigan},
  {Kittiwisit}, {Kohn}, {Kolopanis}, {La Plante}, {Lekalake}, {MacMahon},
  {Malan}, {Malgas}, {Maree}, {Martinot}, {Matsetela}, {Mesinger}, {Molewa},
  {Morales}, {Mosiane}, {Murray}, {Neben}, {Parsons}, {Patra}, {Pieterse},
  {Pober}, {Razavi-Ghods}, {Ringuette}, {Robnett}, {Rosie}, {Sims}, {Smith},
  {Syce}, {Thyagarajan}, {Williams}, \& {Zheng}}]{Kern:2020}
---. 2020{\natexlab{a}}, \apj, 888, 70, \dodoi{10.3847/1538-4357/ab5e8a}

\bibitem[{{Kern} {et~al.}(2020{\natexlab{b}}){Kern}, {Dillon}, {Parsons},
  {Carilli}, {Bernardi}, {Abdurashidova}, {Aguirre}, {Alexander}, {Ali},
  {Balfour}, {Beardsley}, {Billings}, {Bowman}, {Bradley}, {Bull}, {Burba},
  {Carey}, {Cheng}, {DeBoer}, {Dexter}, {de Lera Acedo}, {Ely}, {Ewall-Wice},
  {Fagnoni}, {Fritz}, {Furlanetto}, {Gale-Sides}, {Glendenning}, {Gorthi},
  {Greig}, {Grobbelaar}, {Halday}, {Hazelton}, {Hewitt}, {Hickish}, {Jacobs},
  {Julius}, {Kerrigan}, {Kittiwisit}, {Kohn}, {Kolopanis}, {Lanman}, {La
  Plante}, {Lekalake}, {Liu}, {MacMahon}, {Malan}, {Malgas}, {Maree},
  {Martinot}, {Matsetela}, {Mesinger}, {Molewa}, {Morales}, {Mosiane},
  {Murray}, {Neben}, {Nikolic}, {Nunhokee}, {Patra}, {Pieterse}, {Pober},
  {Razavi-Ghods}, {Ringuette}, {Robnett}, {Rosie}, {Sims}, {Smith}, {Syce},
  {Thyagarajan}, {Williams}, \& {Zheng}}]{Kern:2020b}
{Kern}, N.~S., {Dillon}, J.~S., {Parsons}, A.~R., {et~al.} 2020{\natexlab{b}},
  \apj, 890, 122, \dodoi{10.3847/1538-4357/ab67bc}

\bibitem[{{Kim} {et~al.}(in preparation.)}]{Kim:2021}
{Kim}, H., {et~al.} in preparation.

\bibitem[{{Kingma} \& {Ba}(2014)}]{Kingma:2014}
{Kingma}, D.~P., \& {Ba}, J. 2014, arXiv e-prints, arXiv:1412.6980.
\newblock \doarXiv{1412.6980}

\bibitem[{{Kohn} {et~al.}(2016){Kohn}, {Aguirre}, {Nunhokee}, {Bernardi},
  {Pober}, {Ali}, {Bradley}, {Carilli}, {DeBoer}, {Gugliucci}, {Jacobs},
  {Klima}, {MacMahon}, {Manley}, {Moore}, {Parsons}, {Stefan}, \&
  {Walbrugh}}]{Kohn:2016}
{Kohn}, S.~A., {Aguirre}, J.~E., {Nunhokee}, C.~D., {et~al.} 2016, \apj, 823,
  88, \dodoi{10.3847/0004-637X/823/2/88}

\bibitem[{{Lanman} {et~al.}(2019){Lanman}, {Hazelton}, {Jacobs}, {Kolopanis},
  {Pober}, {Aguirre}, \& {Thyagarajan}}]{Lanman:2019b}
{Lanman}, A., {Hazelton}, B., {Jacobs}, D., {et~al.} 2019, The Journal of Open
  Source Software, 4, 1234, \dodoi{10.21105/joss.01234}

\bibitem[{{Lanman} \& {Pober}(2019)}]{Lanman:2019}
{Lanman}, A.~E., \& {Pober}, J.~C. 2019, \mnras, 487, 5840,
  \dodoi{10.1093/mnras/stz1639}

\bibitem[{{Lenc} {et~al.}(2016){Lenc}, {Gaensler}, {Sun}, {Sadler}, {Willis},
  {Barry}, {Beardsley}, {Bell}, {Bernardi}, {Bowman}, {Briggs}, {Callingham},
  {Cappallo}, {Carroll}, {Corey}, {de Oliveira-Costa}, {Deshpande}, {Dillon},
  {Dwarkanath}, {Emrich}, {Ewall-Wice}, {Feng}, {For}, {Goeke}, {Greenhill},
  {Hancock}, {Hazelton}, {Hewitt}, {Hindson}, {Hurley-Walker},
  {Johnston-Hollitt}, {Jacobs}, {Kapi{\'n}ska}, {Kaplan}, {Kasper}, {Kim},
  {Kratzenberg}, {Line}, {Loeb}, {Lonsdale}, {Lynch}, {McKinley}, {McWhirter},
  {Mitchell}, {Morales}, {Morgan}, {Morgan}, {Murphy}, {Neben}, {Oberoi},
  {Offringa}, {Ord}, {Paul}, {Pindor}, {Pober}, {Prabu}, {Procopio}, {Riding},
  {Rogers}, {Roshi}, {Udaya Shankar}, {Sethi}, {Srivani}, {Staveley-Smith},
  {Subrahmanyan}, {Sullivan}, {Tegmark}, {Thyagarajan}, {Tingay}, {Trott},
  {Waterson}, {Wayth}, {Webster}, {Whitney}, {Williams}, {Williams}, {Wu},
  {Wyithe}, \& {Zheng}}]{Lenc:2016}
{Lenc}, E., {Gaensler}, B.~M., {Sun}, X.~H., {et~al.} 2016, \apj, 830, 38,
  \dodoi{10.3847/0004-637X/830/1/38}

\bibitem[{{Li} {et~al.}(2018){Li}, {Pober}, {Hazelton}, {Barry}, {Morales},
  {Sullivan}, {Parsons}, {Ali}, {Dillon}, {Beardsley}, {Bowman}, {Briggs},
  {Byrne}, {Carroll}, {Crosse}, {Emrich}, {Ewall-Wice}, {Feng}, {Franzen},
  {Hewitt}, {Horsley}, {Jacobs}, {Johnston-Hollitt}, {Jordan}, {Joseph},
  {Kaplan}, {Kenney}, {Kim}, {Kittiwisit}, {Lanman}, {Line}, {McKinley},
  {Mitchell}, {Murray}, {Neben}, {Offringa}, {Pallot}, {Paul}, {Pindor},
  {Procopio}, {Rahimi}, {Riding}, {Sethi}, {Udaya Shankar}, {Steele},
  {Subrahmanian}, {Tegmark}, {Thyagarajan}, {Tingay}, {Trott}, {Walker},
  {Wayth}, {Webster}, {Williams}, {Wu}, \& {Wyithe}}]{Li:2018}
{Li}, W., {Pober}, J.~C., {Hazelton}, B.~J., {et~al.} 2018, \apj, 863, 170,
  \dodoi{10.3847/1538-4357/aad3c3}

\bibitem[{{Li} {et~al.}(2019){Li}, {Pober}, {Barry}, {Hazelton}, {Morales},
  {Trott}, {Lanman}, {Wilensky}, {Sullivan}, {Beardsley}, {Booler}, {Bowman},
  {Byrne}, {Crosse}, {Emrich}, {Franzen}, {Hasegawa}, {Horsley},
  {Johnston-Hollitt}, {Jacobs}, {Jordan}, {Joseph}, {Kaneuji}, {Kaplan},
  {Kenney}, {Kubota}, {Line}, {Lynch}, {McKinley}, {Mitchell}, {Murray},
  {Pallot}, {Pindor}, {Rahimi}, {Riding}, {Sleap}, {Steele}, {Takahashi},
  {Tingay}, {Walker}, {Wayth}, {Webster}, {Williams}, {Wu}, {Wyithe},
  {Yoshiura}, \& {Zheng}}]{Li:2019}
{Li}, W., {Pober}, J.~C., {Barry}, N., {et~al.} 2019, \apj, 887, 141,
  \dodoi{10.3847/1538-4357/ab55e4}

\bibitem[{{Line} {et~al.}(2018){Line}, {McKinley}, {Rasti}, {Bhardwaj},
  {Wayth}, {Webster}, {Ung}, {Emrich}, {Horsley}, {Beardsley}, {Crosse},
  {Franzen}, {Gaensler}, {Johnston-Hollitt}, {Kaplan}, {Kenney}, {Morales},
  {Pallot}, {Steele}, {Tingay}, {Trott}, {Walker}, {Williams}, \&
  {Wu}}]{Line:2018}
{Line}, J.~L.~B., {McKinley}, B., {Rasti}, J., {et~al.} 2018, \pasa, 35, e045,
  \dodoi{10.1017/pasa.2018.30}

\bibitem[{{Liu} \& {Shaw}(2020)}]{Liu:2020}
{Liu}, A., \& {Shaw}, J.~R. 2020, \pasp, 132, 062001,
  \dodoi{10.1088/1538-3873/ab5bfd}

\bibitem[{{Liu} {et~al.}(2010){Liu}, {Tegmark}, {Morrison}, {Lutomirski}, \&
  {Zaldarriaga}}]{Liu:2010}
{Liu}, A., {Tegmark}, M., {Morrison}, S., {Lutomirski}, A., \& {Zaldarriaga},
  M. 2010, \mnras, 408, 1029, \dodoi{10.1111/j.1365-2966.2010.17174.x}

\bibitem[{{McQuinn} {et~al.}(2006){McQuinn}, {Zahn}, {Zaldarriaga},
  {Hernquist}, \& {Furlanetto}}]{McQuinn:2006}
{McQuinn}, M., {Zahn}, O., {Zaldarriaga}, M., {Hernquist}, L., \& {Furlanetto},
  S.~R. 2006, \apj, 653, 815, \dodoi{10.1086/505167}

\bibitem[{{Mertens} {et~al.}(2020){Mertens}, {Mevius}, {Koopmans}, {Offringa},
  {Mellema}, {Zaroubi}, {Brentjens}, {Gan}, {Gehlot}, {Pandey}, {Sardarabadi},
  {Vedantham}, {Yatawatta}, {Asad}, {Ciardi}, {Chapman}, {Gazagnes}, {Ghara},
  {Ghosh}, {Giri}, {Iliev}, {Jeli{\'c}}, {Kooistra}, {Mondal}, {Schaye}, \&
  {Silva}}]{Mertens:2020}
{Mertens}, F.~G., {Mevius}, M., {Koopmans}, L.~V.~E., {et~al.} 2020, \mnras,
  493, 1662, \dodoi{10.1093/mnras/staa327}

\bibitem[{{Mitchell} {et~al.}(2008){Mitchell}, {Greenhill}, {Wayth}, {Sault},
  {Lonsdale}, {Cappallo}, {Morales}, \& {Ord}}]{Mitchell:2008}
{Mitchell}, D.~A., {Greenhill}, L.~J., {Wayth}, R.~B., {et~al.} 2008, IEEE
  Journal of Selected Topics in Signal Processing, 2, 707,
  \dodoi{10.1109/JSTSP.2008.2005327}

\bibitem[{{Moore} {et~al.}(2013){Moore}, {Aguirre}, {Parsons}, {Jacobs}, \&
  {Pober}}]{Moore:2013}
{Moore}, D.~F., {Aguirre}, J.~E., {Parsons}, A.~R., {Jacobs}, D.~C., \&
  {Pober}, J.~C. 2013, \apj, 769, 154, \dodoi{10.1088/0004-637X/769/2/154}

\bibitem[{{Morales} {et~al.}(2012){Morales}, {Hazelton}, {Sullivan}, \&
  {Beardsley}}]{Morales:2012}
{Morales}, M.~F., {Hazelton}, B., {Sullivan}, I., \& {Beardsley}, A. 2012,
  \apj, 752, 137, \dodoi{10.1088/0004-637X/752/2/137}

\bibitem[{{Morales} \& {Hewitt}(2004)}]{Morales:2004}
{Morales}, M.~F., \& {Hewitt}, J. 2004, \apj, 615, 7, \dodoi{10.1086/424437}

\bibitem[{{Neben} {et~al.}(2015){Neben}, {Bradley}, {Hewitt}, {Bernardi},
  {Bowman}, {Briggs}, {Cappallo}, {Deshpande}, {Goeke}, {Greenhill},
  {Hazelton}, {Johnston-Hollitt}, {Kaplan}, {Lonsdale}, {McWhirter},
  {Mitchell}, {Morales}, {Morgan}, {Oberoi}, {Ord}, {Prabu}, {Shankar},
  {Srivani}, {Subrahmanyan}, {Tingay}, {Wayth}, {Webster}, {Williams}, \&
  {Williams}}]{Neben:2015}
{Neben}, A.~R., {Bradley}, R.~F., {Hewitt}, J.~N., {et~al.} 2015, Radio
  Science, 50, 614, \dodoi{10.1002/2015RS005678}

\bibitem[{{Neben} {et~al.}(2016){Neben}, {Bradley}, {Hewitt}, {DeBoer},
  {Parsons}, {Aguirre}, {Ali}, {Cheng}, {Ewall-Wice}, {Patra}, {Thyagarajan},
  {Bowman}, {Dickenson}, {Dillon}, {Doolittle}, {Egan}, {Hedrick}, {Jacobs},
  {Kohn}, {Klima}, {Moodley}, {Saliwanchik}, {Schaffner}, {Shelton}, {Taylor},
  {Taylor}, {Tegmark}, {Wirt}, \& {Zheng}}]{Neben:2016}
---. 2016, \apj, 826, 199, \dodoi{10.3847/0004-637X/826/2/199}

\bibitem[{{Newburgh} {et~al.}(2014){Newburgh}, {Addison}, {Amiri}, {Bandura},
  {Bond}, {Connor}, {Cliche}, {Davis}, {Deng}, {Denman}, {Dobbs}, {Fandino},
  {Fong}, {Gibbs}, {Gilbert}, {Griffin}, {Halpern}, {Hanna}, {Hincks},
  {Hinshaw}, {H{\"o}fer}, {Klages}, {Landecker}, {Masui}, {Parra}, {Pen},
  {Peterson}, {Recnik}, {Shaw}, {Sigurdson}, {Sitwell}, {Smecher}, {Smegal},
  {Vanderlinde}, \& {Wiebe}}]{Newburgh:2014}
{Newburgh}, L.~B., {Addison}, G.~E., {Amiri}, M., {et~al.} 2014, in Society of
  Photo-Optical Instrumentation Engineers (SPIE) Conference Series, Vol. 9145,
  Ground-based and Airborne Telescopes V, ed. L.~M. {Stepp}, R.~{Gilmozzi}, \&
  H.~J. {Hall}, 91454V, \dodoi{10.1117/12.2056962}

\bibitem[{{Nunhokee} {et~al.}(2020){Nunhokee}, {Parsons}, {Kern}, {Nikolic},
  {Pober}, {Bernardi}, {Carilli}, {Abdurashidova}, {Aguirre}, {Alexander},
  {Ali}, {Balfour}, {Beardsley}, {Billings}, {Bowman}, {Bradley}, {Burba},
  {Cheng}, {DeBoer}, {Dexter}, {de Lera Acedo}, {Dillon}, {Ewall-Wice},
  {Fagnoni}, {Fritz}, {Furlanetto}, {Gale-Sides}, {Glendenning}, {Gorthi},
  {Greig}, {Grobbelaar}, {Halday}, {Hazelton}, {Hewitt}, {Jacobs}, {Julius},
  {Kerrigan}, {Kittiwisit}, {Kohn}, {Kolopanis}, {Lanman}, {L Plante},
  {Lekalake}, {Liu}, {MacMahon}, {Malan}, {Malgas}, {Maree}, {Martinot},
  {Matsetela}, {Mesinger}, {Molewa}, {Morales}, {Mosiane}, {Neben}, {Patra},
  {Pieterse}, {Razavi-Ghods}, {Ringuette}, {Robnett}, {Rosie}, {Sims}, {Smith},
  {Syce}, {Thyagarajan}, {Williams}, \& {Zheng}}]{Nunhokee:2020}
{Nunhokee}, C.~D., {Parsons}, A.~R., {Kern}, N.~S., {et~al.} 2020, \apj, 897,
  5, \dodoi{10.3847/1538-4357/ab9634}

\bibitem[{{Offringa} {et~al.}(2019){Offringa}, {Mertens}, \&
  {Koopmans}}]{Offringa:2019}
{Offringa}, A.~R., {Mertens}, F., \& {Koopmans}, L.~V.~E. 2019, \mnras, 484,
  2866, \dodoi{10.1093/mnras/stz175}

\bibitem[{{Orosz} {et~al.}(2019){Orosz}, {Dillon}, {Ewall-Wice}, {Parsons}, \&
  {Thyagarajan}}]{Orosz:2019}
{Orosz}, N., {Dillon}, J.~S., {Ewall-Wice}, A., {Parsons}, A.~R., \&
  {Thyagarajan}, N. 2019, \mnras, 487, 537, \dodoi{10.1093/mnras/stz1287}

\bibitem[{{Parsons} {et~al.}(2012{\natexlab{a}}){Parsons}, {Pober}, {McQuinn},
  {Jacobs}, \& {Aguirre}}]{Parsons:2012}
{Parsons}, A., {Pober}, J., {McQuinn}, M., {Jacobs}, D., \& {Aguirre}, J.
  2012{\natexlab{a}}, \apj, 753, 81, \dodoi{10.1088/0004-637X/753/1/81}

\bibitem[{{Parsons} {et~al.}(2016){Parsons}, {Liu}, {Ali}, \&
  {Cheng}}]{Parsons:2016}
{Parsons}, A.~R., {Liu}, A., {Ali}, Z.~S., \& {Cheng}, C. 2016, \apj, 820, 51,
  \dodoi{10.3847/0004-637X/820/1/51}

\bibitem[{{Parsons} {et~al.}(2012{\natexlab{b}}){Parsons}, {Pober}, {Aguirre},
  {Carilli}, {Jacobs}, \& {Moore}}]{Parsons:2012a}
{Parsons}, A.~R., {Pober}, J.~C., {Aguirre}, J.~E., {et~al.}
  2012{\natexlab{b}}, \apj, 756, 165, \dodoi{10.1088/0004-637X/756/2/165}

\bibitem[{{Parsons} {et~al.}(2014){Parsons}, {Liu}, {Aguirre}, {Ali},
  {Bradley}, {Carilli}, {DeBoer}, {Dexter}, {Gugliucci}, {Jacobs}, {Klima},
  {MacMahon}, {Manley}, {Moore}, {Pober}, {Stefan}, \&
  {Walbrugh}}]{Parsons:2014}
{Parsons}, A.~R., {Liu}, A., {Aguirre}, J.~E., {et~al.} 2014, \apj, 788, 106,
  \dodoi{10.1088/0004-637X/788/2/106}

\bibitem[{{Patil} {et~al.}(2016){Patil}, {Yatawatta}, {Zaroubi}, {Koopmans},
  {de Bruyn}, {Jeli{\'c}}, {Ciardi}, {Iliev}, {Mevius}, {Pandey}, \&
  {Gehlot}}]{Patil:2016}
{Patil}, A.~H., {Yatawatta}, S., {Zaroubi}, S., {et~al.} 2016, \mnras, 463,
  4317, \dodoi{10.1093/mnras/stw2277}

\bibitem[{{Patil} {et~al.}(2017){Patil}, {Yatawatta}, {Koopmans}, {de Bruyn},
  {Brentjens}, {Zaroubi}, {Asad}, {Hatef}, {Jeli{\'c}}, {Mevius}, {Offringa},
  {Pandey}, {Vedantham}, {Abdalla}, {Brouw}, {Chapman}, {Ciardi}, {Gehlot},
  {Ghosh}, {Harker}, {Iliev}, {Kakiichi}, {Majumdar}, {Mellema}, {Silva},
  {Schaye}, {Vrbanec}, \& {Wijnholds}}]{Patil:2017}
{Patil}, A.~H., {Yatawatta}, S., {Koopmans}, L.~V.~E., {et~al.} 2017, \apj,
  838, 65, \dodoi{10.3847/1538-4357/aa63e7}

\bibitem[{{Patra} {et~al.}(2018){Patra}, {Parsons}, {DeBoer}, {Thyagarajan},
  {Ewall-Wice}, {Hsyu}, {Leung}, {Day}, {de Lera Acedo}, {Aguirre},
  {Alexander}, {Ali}, {Beardsley}, {Bowman}, {Bradley}, {Carilli}, {Cheng},
  {Dillon}, {Fadana}, {Fagnoni}, {Fritz}, {Furlanetto}, {Glendenning}, {Greig},
  {Grobbelaar}, {Hazelton}, {Jacobs}, {Julius}, {Kariseb}, {Kohn}, {Lebedeva},
  {Lekalake}, {Liu}, {Loots}, {MacMahon}, {Malan}, {Malgas}, {Maree},
  {Martinot}, {Mathison}, {Matsetela}, {Mesinger}, {Morales}, {Neben},
  {Pieterse}, {Pober}, {Razavi-Ghods}, {Ringuette}, {Robnett}, {Rosie}, {Sell},
  {Smith}, {Syce}, {Tegmark}, {Williams}, \& {Zheng}}]{Patra:2018}
{Patra}, N., {Parsons}, A.~R., {DeBoer}, D.~R., {et~al.} 2018, Experimental
  Astronomy, 45, 177, \dodoi{10.1007/s10686-017-9563-0}

\bibitem[{{Pober} {et~al.}(2012){Pober}, {Parsons}, {Jacobs}, {Aguirre},
  {Bradley}, {Carilli}, {Gugliucci}, {Moore}, \& {Parashare}}]{Pober:2012}
{Pober}, J.~C., {Parsons}, A.~R., {Jacobs}, D.~C., {et~al.} 2012, \aj, 143, 53,
  \dodoi{10.1088/0004-6256/143/2/53}

\bibitem[{{Pober} {et~al.}(2013){Pober}, {Parsons}, {Aguirre}, {Ali},
  {Bradley}, {Carilli}, {DeBoer}, {Dexter}, {Gugliucci}, {Jacobs}, {Klima},
  {MacMahon}, {Manley}, {Moore}, {Stefan}, \& {Walbrugh}}]{Pober:2013}
{Pober}, J.~C., {Parsons}, A.~R., {Aguirre}, J.~E., {et~al.} 2013, \apjl, 768,
  L36, \dodoi{10.1088/2041-8205/768/2/L36}

\bibitem[{{Pober} {et~al.}(2014){Pober}, {Liu}, {Dillon}, {Aguirre}, {Bowman},
  {Bradley}, {Carilli}, {DeBoer}, {Hewitt}, {Jacobs}, {McQuinn}, {Morales},
  {Parsons}, {Tegmark}, \& {Werthimer}}]{Pober:2014}
{Pober}, J.~C., {Liu}, A., {Dillon}, J.~S., {et~al.} 2014, \apj, 782, 66,
  \dodoi{10.1088/0004-637X/782/2/66}

\bibitem[{{Prabu} {et~al.}(2015){Prabu}, {Srivani}, {Roshi}, {Kamini},
  {Madhavi}, {Emrich}, {Crosse}, {Williams}, {Waterson}, {Deshpande},
  {Shankar}, {Subrahmanyan}, {Briggs}, {Goeke}, {Tingay}, {Johnston-Hollitt},
  {R}, {Morgan}, {Pathikulangara}, {Bunton}, {Hampson}, {Williams}, {Ord},
  {Wayth}, {Kumar}, {Morales}, {deSouza}, {Kratzenberg}, {Pallot}, {McWhirter},
  {Hazelton}, {Arcus}, {Barnes}, {Bernardi}, {Booler}, {Bowman}, {Cappallo},
  {Corey}, {Greenhill}, {Herne}, {Hewitt}, {Kaplan}, {Kasper}, {Kincaid},
  {Koenig}, {Lonsdale}, {Lynch}, {Mitchell}, {Oberoi}, {Remillard}, {Rogers},
  {Salah}, {Sault}, {Stevens}, {Tremblay}, {Webster}, {Whitney}, \&
  {Wyithe}}]{Prabu:2015}
{Prabu}, T., {Srivani}, K.~S., {Roshi}, D.~A., {et~al.} 2015, Experimental
  Astronomy, 39, 73, \dodoi{10.1007/s10686-015-9444-3}

\bibitem[{{Saliwanchik} {et~al.}(2021){Saliwanchik}, {Ewall-Wice}, {Chichton},
  {Kuhn}, {{\~A}-l{\c{c}}ek}, {Bandura}, {Bucher}, {Chang}, {Chiang},
  {Gerodias}, {Kesebonye}, {MacKay}, {Moodley}, {Newburgh}, {Nistane},
  {Peterson}, {Pieters}, {Pieterse}, {Vanderlinde}, {Sievers}, {Weltman}, \&
  {Wulf}}]{Saliwanchik:2018}
{Saliwanchik}, B. R.~B., {Ewall-Wice}, A., {Chichton}, D., {et~al.} 2021, in
  Society of Photo-Optical Instrumentation Engineers (SPIE) Conference Series,
  Vol. 11445, Society of Photo-Optical Instrumentation Engineers (SPIE)
  Conference Series, 114455O, \dodoi{10.1117/12.2552508}

\bibitem[{{Salvini} \& {Wijnholds}(2014)}]{Salvini:2014}
{Salvini}, S., \& {Wijnholds}, S.~J. 2014, \aap, 571, A97,
  \dodoi{10.1051/0004-6361/201424487}

\bibitem[{{Sievers}(2017)}]{Sievers:2017}
{Sievers}, J.~L. 2017, arXiv e-prints, arXiv:1701.01860.
\newblock \doarXiv{1701.01860}

\bibitem[{{Slepian}(1978)}]{Slepian:1978}
{Slepian}, D. 1978, AT T Technical Journal, 57, 1371

\bibitem[{{Sob} {et~al.}(2020){Sob}, {Bester}, {Smirnov}, {Kenyon}, \&
  {Grobler}}]{Sob:2020}
{Sob}, U.~M., {Bester}, H.~L., {Smirnov}, O.~M., {Kenyon}, J.~S., \& {Grobler},
  T.~L. 2020, \mnras, 491, 1026, \dodoi{10.1093/mnras/stz3037}

\bibitem[{{Sokolowski} {et~al.}(2017){Sokolowski}, {Colegate}, {Sutinjo},
  {Ung}, {Wayth}, {Hurley-Walker}, {Lenc}, {Pindor}, {Morgan}, {Kaplan},
  {Bell}, {Callingham}, {Dwarakanath}, {For}, {Gaensler}, {Hancock}, {Hindson},
  {Johnston-Hollitt}, {Kapi{\'n}ska}, {McKinley}, {Offringa}, {Procopio},
  {Staveley-Smith}, {Wu}, \& {Zheng}}]{Sokolowski:2017}
{Sokolowski}, M., {Colegate}, T., {Sutinjo}, A.~T., {et~al.} 2017, \pasa, 34,
  e062, \dodoi{10.1017/pasa.2017.54}

\bibitem[{{Subrahmanya} {et~al.}(2017){Subrahmanya}, {Manoharan}, \&
  {Chengalur}}]{Subrahmanya:2017}
{Subrahmanya}, C.~R., {Manoharan}, P.~K., \& {Chengalur}, J.~N. 2017, Journal
  of Astrophysics and Astronomy, 38, 10, \dodoi{10.1007/s12036-017-9430-4}

\bibitem[{{Sutinjo} {et~al.}(2015){Sutinjo}, {Colegate}, {Wayth}, {Hall}, {de
  Lera Acedo}, {Booler}, {Faulkner}, {Feng}, {Hurley-Walker}, {Juswardy},
  {Padhi}, {Razavi-Ghods}, {Sokolowski}, {Tingay}, \& {Bij de
  Vaate}}]{Sutinjo:2015}
{Sutinjo}, A.~T., {Colegate}, T.~M., {Wayth}, R.~B., {et~al.} 2015, IEEE
  Transactions on Antennas and Propagation, 63, 5433,
  \dodoi{10.1109/TAP.2015.2487504}

\bibitem[{{Tasse} {et~al.}(2018){Tasse}, {Hugo}, {Mirmont}, {Smirnov},
  {Atemkeng}, {Bester}, {Hardcastle}, {Lakhoo}, {Perkins}, \&
  {Shimwell}}]{Tasse:2018}
{Tasse}, C., {Hugo}, B., {Mirmont}, M., {et~al.} 2018, \aap, 611, A87,
  \dodoi{10.1051/0004-6361/201731474}

\bibitem[{{The HERA Collaboration} {et~al.}(2021){The HERA Collaboration},
  {Abdurashidova}, {Aguirre}, {Alexander}, {Ali}, {Balfour}, {Beardsley},
  {Bernardi}, {Billings}, {Bowman}, {Bradley}, {Bull}, {Burba}, {Carey},
  {Carilli}, {Cheng}, {DeBoer}, {Dexter}, {de Lera Acedo}, {Dibblee-Barkman},
  {Dillon}, {Ely}, {Ewall-Wice}, {Fagnoni}, {Fritz}, {Furlanetto},
  {Gale-Sides}, {Glendenning}, {Gorthi}, {Greig}, {Grobbelaar}, {Halday},
  {Hazelton}, {Hewitt}, {Hickish}, {Jacobs}, {Julius}, {Kern}, {Kerrigan},
  {Kittiwisit}, {Kohn}, {Kolopanis}, {Lanman}, {La Plante}, {Lekalake},
  {Lewis}, {Liu}, {MacMahon}, {Malan}, {Malgas}, {Maree}, {Martinot},
  {Matsetela}, {Mesinger}, {Molewa}, {Morales}, {Mosiane}, {Murray}, {Neben},
  {Nikolic}, {Nunhokee}, {Parsons}, {Patra}, {Pascua}, {Pieterse}, {Pober},
  {Razavi-Ghods}, {Ringuette}, {Robnett}, {Rosie}, {Sims}, {Singh}, {Smith},
  {Syce}, {Thyagarajan}, {Williams}, \& {Zheng}}]{HERA:2021}
{The HERA Collaboration}, {Abdurashidova}, Z., {Aguirre}, J.~E., {et~al.} 2021,
  arXiv e-prints, arXiv:2108.02263.
\newblock \doarXiv{2108.02263}

\bibitem[{{Thyagarajan} {et~al.}(2016){Thyagarajan}, {Parsons}, {DeBoer},
  {Bowman}, {Ewall-Wice}, {Neben}, \& {Patra}}]{Thyagarajan:2016}
{Thyagarajan}, N., {Parsons}, A.~R., {DeBoer}, D.~R., {et~al.} 2016, \apj, 825,
  9, \dodoi{10.3847/0004-637X/825/1/9}

\bibitem[{{Thyagarajan} {et~al.}(2013){Thyagarajan}, {Udaya Shankar},
  {Subrahmanyan}, {Arcus}, {Bernardi}, {Bowman}, {Briggs}, {Bunton},
  {Cappallo}, {Corey}, {deSouza}, {Emrich}, {Gaensler}, {Goeke}, {Greenhill},
  {Hazelton}, {Herne}, {Hewitt}, {Johnston-Hollitt}, {Kaplan}, {Kasper},
  {Kincaid}, {Koenig}, {Kratzenberg}, {Lonsdale}, {Lynch}, {McWhirter},
  {Mitchell}, {Morales}, {Morgan}, {Oberoi}, {Ord}, {Pathikulangara},
  {Remillard}, {Rogers}, {Anish Roshi}, {Salah}, {Sault}, {Srivani}, {Stevens},
  {Thiagaraj}, {Tingay}, {Wayth}, {Waterson}, {Webster}, {Whitney}, {Williams},
  {Williams}, \& {Wyithe}}]{Thyagarajan:2013}
{Thyagarajan}, N., {Udaya Shankar}, N., {Subrahmanyan}, R., {et~al.} 2013,
  \apj, 776, 6, \dodoi{10.1088/0004-637X/776/1/6}

\bibitem[{{Tingay} {et~al.}(2013){Tingay}, {Goeke}, {Bowman}, {Emrich}, {Ord},
  {Mitchell}, {Morales}, {Booler}, {Crosse}, {Wayth}, {Lonsdale}, {Tremblay},
  {Pallot}, {Colegate}, {Wicenec}, {Kudryavtseva}, {Arcus}, {Barnes},
  {Bernardi}, {Briggs}, {Burns}, {Bunton}, {Cappallo}, {Corey}, {Deshpande},
  {Desouza}, {Gaensler}, {Greenhill}, {Hall}, {Hazelton}, {Herne}, {Hewitt},
  {Johnston-Hollitt}, {Kaplan}, {Kasper}, {Kincaid}, {Koenig}, {Kratzenberg},
  {Lynch}, {Mckinley}, {Mcwhirter}, {Morgan}, {Oberoi}, {Pathikulangara},
  {Prabu}, {Remillard}, {Rogers}, {Roshi}, {Salah}, {Sault}, {Udaya-Shankar},
  {Schlagenhaufer}, {Srivani}, {Stevens}, {Subrahmanyan}, {Waterson},
  {Webster}, {Whitney}, {Williams}, {Williams}, \& {Wyithe}}]{Tingay:2013}
{Tingay}, S.~J., {Goeke}, R., {Bowman}, J.~D., {et~al.} 2013, \pasa, 30, e007,
  \dodoi{10.1017/pasa.2012.007}

\bibitem[{{Trott} \& {Wayth}(2016)}]{Trott:2016}
{Trott}, C.~M., \& {Wayth}, R.~B. 2016, \pasa, 33, e019,
  \dodoi{10.1017/pasa.2016.18}

\bibitem[{{van Haarlem} {et~al.}(2013){van Haarlem}, {Wise}, {Gunst}, {Heald},
  {McKean}, {Hessels}, {de Bruyn}, {Nijboer}, {Swinbank}, {Fallows},
  {Brentjens}, {Nelles}, {Beck}, {Falcke}, {Fender}, {H{\"o}randel},
  {Koopmans}, {Mann}, {Miley}, {R{\"o}ttgering}, {Stappers}, {Wijers},
  {Zaroubi}, {van den Akker}, {Alexov}, {Anderson}, {Anderson}, {van Ardenne},
  {Arts}, {Asgekar}, {Avruch}, {Batejat}, {B{\"a}hren}, {Bell}, {Bell}, {van
  Bemmel}, {Bennema}, {Bentum}, {Bernardi}, {Best}, {B{\^\i}rzan}, {Bonafede},
  {Boonstra}, {Braun}, {Bregman}, {Breitling}, {van de Brink}, {Broderick},
  {Broekema}, {Brouw}, {Br{\"u}ggen}, {Butcher}, {van Cappellen}, {Ciardi},
  {Coenen}, {Conway}, {Coolen}, {Corstanje}, {Damstra}, {Davies}, {Deller},
  {Dettmar}, {van Diepen}, {Dijkstra}, {Donker}, {Doorduin}, {Dromer}, {Drost},
  {van Duin}, {Eisl{\"o}ffel}, {van Enst}, {Ferrari}, {Frieswijk}, {Gankema},
  {Garrett}, {de Gasperin}, {Gerbers}, {de Geus}, {Grie{\ss}meier}, {Grit},
  {Gruppen}, {Hamaker}, {Hassall}, {Hoeft}, {Holties}, {Horneffer}, {van der
  Horst}, {van Houwelingen}, {Huijgen}, {Iacobelli}, {Intema}, {Jackson},
  {Jelic}, {de Jong}, {Juette}, {Kant}, {Karastergiou}, {Koers}, {Kollen},
  {Kondratiev}, {Kooistra}, {Koopman}, {Koster}, {Kuniyoshi}, {Kramer},
  {Kuper}, {Lambropoulos}, {Law}, {van Leeuwen}, {Lemaitre}, {Loose}, {Maat},
  {Macario}, {Markoff}, {Masters}, {McFadden}, {McKay-Bukowski}, {Meijering},
  {Meulman}, {Mevius}, {Middelberg}, {Millenaar}, {Miller-Jones}, {Mohan},
  {Mol}, {Morawietz}, {Morganti}, {Mulcahy}, {Mulder}, {Munk}, {Nieuwenhuis},
  {van Nieuwpoort}, {Noordam}, {Norden}, {Noutsos}, {Offringa}, {Olofsson},
  {Omar}, {Orr{\'u}}, {Overeem}, {Paas}, {Pandey-Pommier}, {Pandey}, {Pizzo},
  {Polatidis}, {Rafferty}, {Rawlings}, {Reich}, {de Reijer}, {Reitsma},
  {Renting}, {Riemers}, {Rol}, {Romein}, {Roosjen}, {Ruiter}, {Scaife}, {van
  der Schaaf}, {Scheers}, {Schellart}, {Schoenmakers}, {Schoonderbeek},
  {Serylak}, {Shulevski}, {Sluman}, {Smirnov}, {Sobey}, {Spreeuw}, {Steinmetz},
  {Sterks}, {Stiepel}, {Stuurwold}, {Tagger}, {Tang}, {Tasse}, {Thomas},
  {Thoudam}, {Toribio}, {van der Tol}, {Usov}, {van Veelen}, {van der Veen},
  {ter Veen}, {Verbiest}, {Vermeulen}, {Vermaas}, {Vocks}, {Vogt}, {de Vos},
  {van der Wal}, {van Weeren}, {Weggemans}, {Weltevrede}, {White}, {Wijnholds},
  {Wilhelmsson}, {Wucknitz}, {Yatawatta}, {Zarka}, {Zensus}, \& {van
  Zwieten}}]{VanHaarlem:2013}
{van Haarlem}, M.~P., {Wise}, M.~W., {Gunst}, A.~W., {et~al.} 2013, \aap, 556,
  A2, \dodoi{10.1051/0004-6361/201220873}

\bibitem[{{van Weeren} {et~al.}(2016){van Weeren}, {Williams}, {Hardcastle},
  {Shimwell}, {Rafferty}, {Sabater}, {Heald}, {Sridhar}, {Dijkema}, {Brunetti},
  {Br{\"u}ggen}, {Andrade-Santos}, {Ogrean}, {R{\"o}ttgering}, {Dawson},
  {Forman}, {de Gasperin}, {Jones}, {Miley}, {Rudnick}, {Sarazin}, {Bonafede},
  {Best}, {B{\^\i}rzan}, {Cassano}, {Chy{\.z}y}, {Croston}, {Ensslin},
  {Ferrari}, {Hoeft}, {Horellou}, {Jarvis}, {Kraft}, {Mevius}, {Intema},
  {Murray}, {Orr{\'u}}, {Pizzo}, {Simionescu}, {Stroe}, {van der Tol}, \&
  {White}}]{VanWeeren:2016}
{van Weeren}, R.~J., {Williams}, W.~L., {Hardcastle}, M.~J., {et~al.} 2016,
  \apjs, 223, 2, \dodoi{10.3847/0067-0049/223/1/2}

\bibitem[{{Vedantham} {et~al.}(2012){Vedantham}, {Udaya Shankar}, \&
  {Subrahmanyan}}]{Vedantham:2012}
{Vedantham}, H., {Udaya Shankar}, N., \& {Subrahmanyan}, R. 2012, \apj, 745,
  176, \dodoi{10.1088/0004-637X/745/2/176}

\bibitem[{{Vedantham} \& {Koopmans}(2015)}]{Vedantham:2015}
{Vedantham}, H.~K., \& {Koopmans}, L.~V.~E. 2015, \mnras, 453, 925,
  \dodoi{10.1093/mnras/stv1594}

\bibitem[{Virtanen {et~al.}(2020)Virtanen, Gommers, Oliphant, Haberland, Reddy,
  Cournapeau, Burovski, Peterson, Weckesser, Bright, {van der Walt}, Brett,
  Wilson, Millman, Mayorov, Nelson, Jones, Kern, Larson, Carey, Polat, Feng,
  Moore, {VanderPlas}, Laxalde, Perktold, Cimrman, Henriksen, Quintero, Harris,
  Archibald, Ribeiro, Pedregosa, {van Mulbregt}, \& {SciPy 1.0
  Contributors}}]{Scipy:2020}
Virtanen, P., Gommers, R., Oliphant, T.~E., {et~al.} 2020, Nature Methods, 17,
  261, \dodoi{10.1038/s41592-019-0686-2}

\bibitem[{{Wayth} {et~al.}(2018){Wayth}, {Tingay}, {Trott}, {Emrich},
  {Johnston-Hollitt}, {McKinley}, {Gaensler}, {Beardsley}, {Booler}, {Crosse},
  {Franzen}, {Horsley}, {Kaplan}, {Kenney}, {Morales}, {Pallot}, {Sleap},
  {Steele}, {Walker}, {Williams}, {Wu}, {Cairns}, {Filipovic}, {Johnston},
  {Murphy}, {Quinn}, {Staveley-Smith}, {Webster}, \& {Wyithe}}]{Wayth:2018}
{Wayth}, R.~B., {Tingay}, S.~J., {Trott}, C.~M., {et~al.} 2018, \pasa, 35,
  e033, \dodoi{10.1017/pasa.2018.37}

\bibitem[{{Wieringa}(1992)}]{Wieringa:1992}
{Wieringa}, M.~H. 1992, Experimental Astronomy, 2, 203,
  \dodoi{10.1007/BF00420576}

\bibitem[{{Yatawatta}(2016)}]{Yatawatta:2016}
{Yatawatta}, S. 2016, arXiv e-prints, arXiv:1605.09219.
\newblock \doarXiv{1605.09219}

\bibitem[{{Yoshiura} {et~al.}(2021){Yoshiura}, {Pindor}, {Line}, {Barry},
  {Trott}, {Beardsley}, {Bowman}, {Byrne}, {Chokshi}, {Hazelton}, {Hasegawa},
  {Howard}, {Greig}, {Jacobs}, {Jordan}, {Joseph}, {Kolopanis}, {Lynch},
  {McKinley}, {Mitchell}, {Morales}, {Murray}, {Pober}, {Rahimi}, {Takahashi},
  {Tingay}, {Wayth}, {Webster}, {Wilensky}, {Wyithe}, {Zhang}, \&
  {Zheng}}]{Yoshiura:2021}
{Yoshiura}, S., {Pindor}, B., {Line}, J.~L.~B., {et~al.} 2021, \mnras, 505,
  4775, \dodoi{10.1093/mnras/stab1560}

\bibitem[{{Zarka} {et~al.}(2012){Zarka}, {Girard}, {Tagger}, \&
  {Denis}}]{Zarka:2012}
{Zarka}, P., {Girard}, J.~N., {Tagger}, M., \& {Denis}, L. 2012, in SF2A-2012:
  Proceedings of the Annual meeting of the French Society of Astronomy and
  Astrophysics, ed. S.~{Boissier}, P.~{de Laverny}, N.~{Nardetto}, R.~{Samadi},
  D.~{Valls-Gabaud}, \& H.~{Wozniak}, 687--694

\bibitem[{{Zheng} {et~al.}(2013){Zheng}, {Tegmark}, {Buza}, {Dillon},
  {Gharibyan}, {Hickish}, {Kunz}, {Liu}, {Losh}, {Lutomirski}, {Morrison},
  {Narayanan}, {Perko}, {Rosner}, {Sanchez}, {Schutz}, {Tribiano},
  {Zaldarriaga}, {Zarb Adami}, {Zelko}, {Zheng}, {Armstrong}, {Bradley},
  {Dexter}, {Ewall-Wice}, {Magro}, {Matejek}, {Morgan}, {Neben}, {Pan},
  {Peterson}, {Su}, {Villasenor}, {Williams}, {Yang}, \& {Zhu}}]{Zheng:2013}
{Zheng}, H., {Tegmark}, M., {Buza}, V., {et~al.} 2013, arXiv e-prints,
  arXiv:1309.2639.
\newblock \doarXiv{1309.2639}

\bibitem[{{Zheng} {et~al.}(2014){Zheng}, {Tegmark}, {Buza}, {Dillon},
  {Gharibyan}, {Hickish}, {Kunz}, {Liu}, {Losh}, {Lutomirski}, {Morrison},
  {Narayanan}, {Perko}, {Rosner}, {Sanchez}, {Schutz}, {Tribiano}, {Valdez},
  {Yang}, {Adami}, {Zelko}, {Zheng}, {Armstrong}, {Bradley}, {Dexter},
  {Ewall-Wice}, {Magro}, {Matejek}, {Morgan}, {Neben}, {Pan}, {Penna},
  {Peterson}, {Su}, {Villasenor}, {Williams}, \& {Zhu}}]{Zheng:2014}
---. 2014, \mnras, 445, 1084, \dodoi{10.1093/mnras/stu1773}

\end{thebibliography}
\bibliographystyle{aasjournal}



\end{document}